\documentclass[preprint,prd,amsmath,amssymb,superscriptaddress,nofootinbib]{revtex4-1}
\usepackage[dvips]{graphicx}
\usepackage{amssymb}
\usepackage{rotating}
\usepackage{mathtools}
\usepackage{color}

\allowdisplaybreaks

\newcommand\op[1]{\hat{#1}}						  
\newcommand\vecl[1]{{\bf{#1}}}         		  
\newcommand\vecs[1]{\boldsymbol{#1}}     		  
\newcommand\sqb[1]{\left[#1\right]}     		  
\newcommand\nb[1]{(#1)}                 		  
\newcommand\nba[1]{\left(#1\right)}     		  
\newcommand{\ket}[1]{\left| #1 \right>} 		  
\newcommand{\bra}[1]{\left< #1 \right|}  		  
\newcommand{\abs}[1]{\left| #1 \right|}		  
\newcommand\nn[0]{\nonumber}					  
\newcommand\ovbb[0]{0\nu\beta\beta}			  
\newcommand\nme[0]{\mathcal{M}}				  
\newcommand{\signs}[3]{(#1\ #2\ #3)\ }			  
\newcommand{\foursigns}[4]{(#1\ #2\ #3 \ #4)\ }  

\begin{document}

\title{Short-Range Neutrinoless Double Beta Decay Mechanisms} 

\author{Lukas Graf}
\email{lukas.graf.14@ucl.ac.uk}
\affiliation{Department of Physics and Astronomy, University College London, Gower Street, London WC1E 6BT, UK}
\affiliation{Center for Theoretical Physics, Sloane Physics Laboratory, Yale University, New Haven, Connecticut 06520-8120, USA}
\affiliation{Perimeter Institute for Theoretical Physics, Caroline Street North, Waterloo, Ontario N2L 2Y5, Canada}

\author{Frank F. Deppisch}
\email{f.deppisch@ucl.ac.uk}
\affiliation{Department of Physics and Astronomy, University College London, Gower Street, London WC1E 6BT, UK}

\author{Francesco Iachello}
\email{francesco.iachello@yale.edu}
\affiliation{Center for Theoretical Physics, Sloane Physics Laboratory, Yale University, New Haven, Connecticut 06520-8120, USA}

\author{Jenni Kotila}
\email{jenni.kotila@jyu.fi}
\affiliation{Finnish Institute for Educational Research, University of Jyv{\"a}skyl{\"a}, P.O. Box 35, FI-40014 University of Jyv{\"a}skyl{\"a}, Finland}

\begin{abstract}
	Neutrinoless double beta decay can significantly help to shed light on the issue of non-zero neutrino mass, as observation of this lepton number violating process would imply neutrinos are Majorana particles. However, the underlying interaction does not have to be as simple as the standard neutrino mass mechanism. The entire variety of neutrinoless double beta decay mechanisms can be approached effectively. In this work we focus on a theoretical description of short-range effective contributions to neutrinoless double beta decay, which are equivalent to 9-dimensional effective operators incorporating the appropriate field content. We give a detailed derivation of the nuclear matrix elements and phase space factors corresponding to individual terms of the effective Lagrangian. Using these, we provide general formulae for the neutrinoless double beta decay half-life and angular correlation of the outgoing electrons.
\end{abstract}

\maketitle

\section{Introduction}
\label{sec:intro}

While the Standard Model (SM) gauge group $SU(3)_C \times SU(2)_L \times U(1)_Y$ perfectly explains the interactions we observe, its breaking also provides masses to the charged fermions via the Higgs mechanism. The discovery of the Higgs at the Large Hadron Collider (LHC)~\cite{Aad:2012tfa, Chatrchyan:2012ufa} allows us to probe and test this mass mechanism in the SM. Yet, neutrinos continue to evade our understanding as only left-handed neutrinos exist in the SM and they therefore cannot acquire a so-called Dirac mass like the other SM fermions. Neutrino oscillation experiments \cite{Agashe:2014kda} have unambiguously shown though that at least two of the three known neutrino species have finite masses and while they are not sensitive to the absolute neutrino masses, they point to mass scales of order $10^{-2}$~eV to $5\times 10^{-2}$~eV. In addition, cosmological observations set an upper limit on the sum of neutrino masses $\Sigma m_\nu \lesssim 0.15$~eV \cite{Ade:2015xua}, assuming the standard cosmological model and with the exact value depending on the observational data considered.

Neutrinos could be of Dirac type as the other SM fermions, but this requires a new right-handed neutrino $\nu_R$ and tiny Yukawa couplings $\lesssim 10^{-12}$, which is rather unnatural. Because the right-handed neutrinos would be completely sterile with respect to the SM gauge interactions, it is on the other hand theoretically indicated that they acquire a Majorana mass $M$ of the form $M \bar \nu_L C \bar\nu_L^T$. It is generically expected to be of the order of a large new physics scale $\Lambda_\text{NP} \approx M$ associated with the breaking of lepton number $L$ symmetry. Via the Yukawa couplings between left and right-handed neutrinos, it will induce an effective dimension-5 operator, $\Lambda^{-1}_\text{NP}(L L H H)$ \cite{Weinberg:1979sa}, where $L$ and $H$ represent the $SU(2)_L$ left-handed lepton and the Higgs doublets, respectively. After electroweak (EW) symmetry breaking, a small effective Majorana mass $m_\nu \sim m_\text{EW}^2 / \Lambda_\text{NP}$ is generated for the active neutrinos. This corresponds to the famous seesaw mechanism \cite{Minkowski:1977sc, mohapatra:1979ia, Yanagida:1979as, seesaw:1979, Schechter:1980gr}, with a scale $\Lambda_\text{NP}$ naturally of the order $10^{14}$~GeV to explain the light neutrino masses $m_\nu \approx 0.1$~eV. 

While the most prominent scenario, the high-scale seesaw mechanism is not the only possibility to generate light neutrino masses; there are numerous other ways by incorporating lepton number violation at low scales in secluded sectors, at higher loop order and when allowing higher-dimensional effective interactions beyond the Weinberg operator. If the $L$ breaking occurs closer to the EW scale, higher-dimensional $L$-breaking operators will be important for phenomenology, and specifically they will potentially induce neutrinoless double beta $\ovbb$ decay.

The search for $\ovbb$ decay is the most sensitive approach to probe Majorana neutrino masses. The experimentally most stringent lower limit on the decay half life $T_{1/2}$ is derived using the Xenon isotope ${}^{136}_{\phantom{1}54}$Xe,
\begin{align}
	T_{1/2}^\text{Xe} \equiv T_{1/2}\left({}^{136}_{\phantom{1}54}\text{Xe} \to {}^{136}_{\phantom{1}56}\text{Ba} + e^- e^-\right) \gtrsim 10^{26}~\text{y}.
\end{align}
However, Majorana neutrino masses are not the only element of Beyond-the-SM (BSM) physics which can induce it. As hinted on above, other mechanisms of $\ovbb$ decay where the LNV originates from LNV masses and couplings of new particles appearing in various possible extensions of the SM. The same couplings and states will also induce light neutrino masses due to the Schechter-Valle black box argument \cite{Schechter:1981bd}, but the resulting contribution will not be necessarily dominant. Instead, we consider the $\ovbb$ decay rate by expressing high scale new physics contributions in terms of effective low-energy operators \cite{Pas:1999fc, Pas:2000vn, delAguila:2011gr, delAguila:2012nu}.
\begin{figure}[t!]
	\centering
	\includegraphics[clip,width=0.32\textwidth]{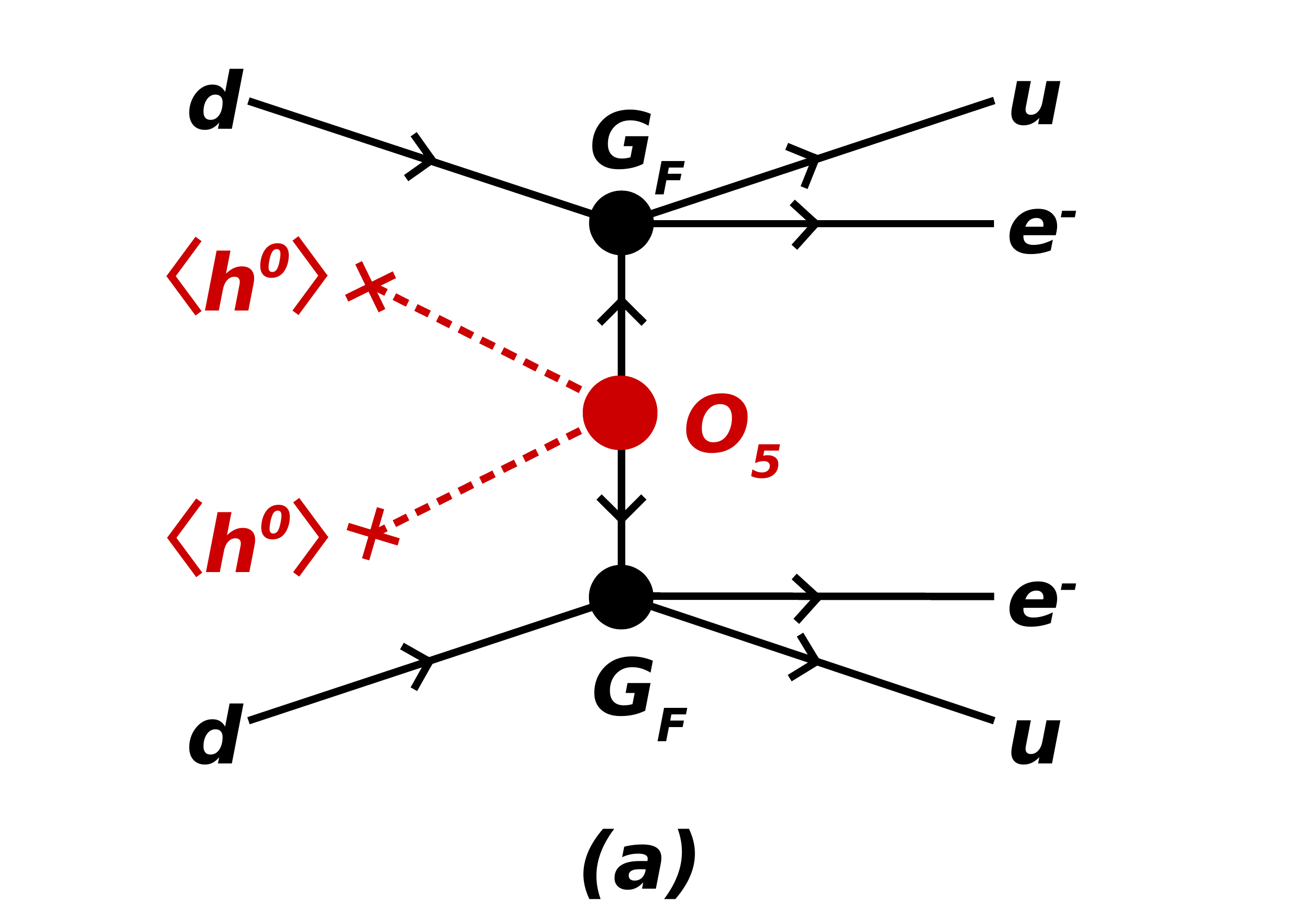}
	\includegraphics[clip,width=0.32\textwidth]{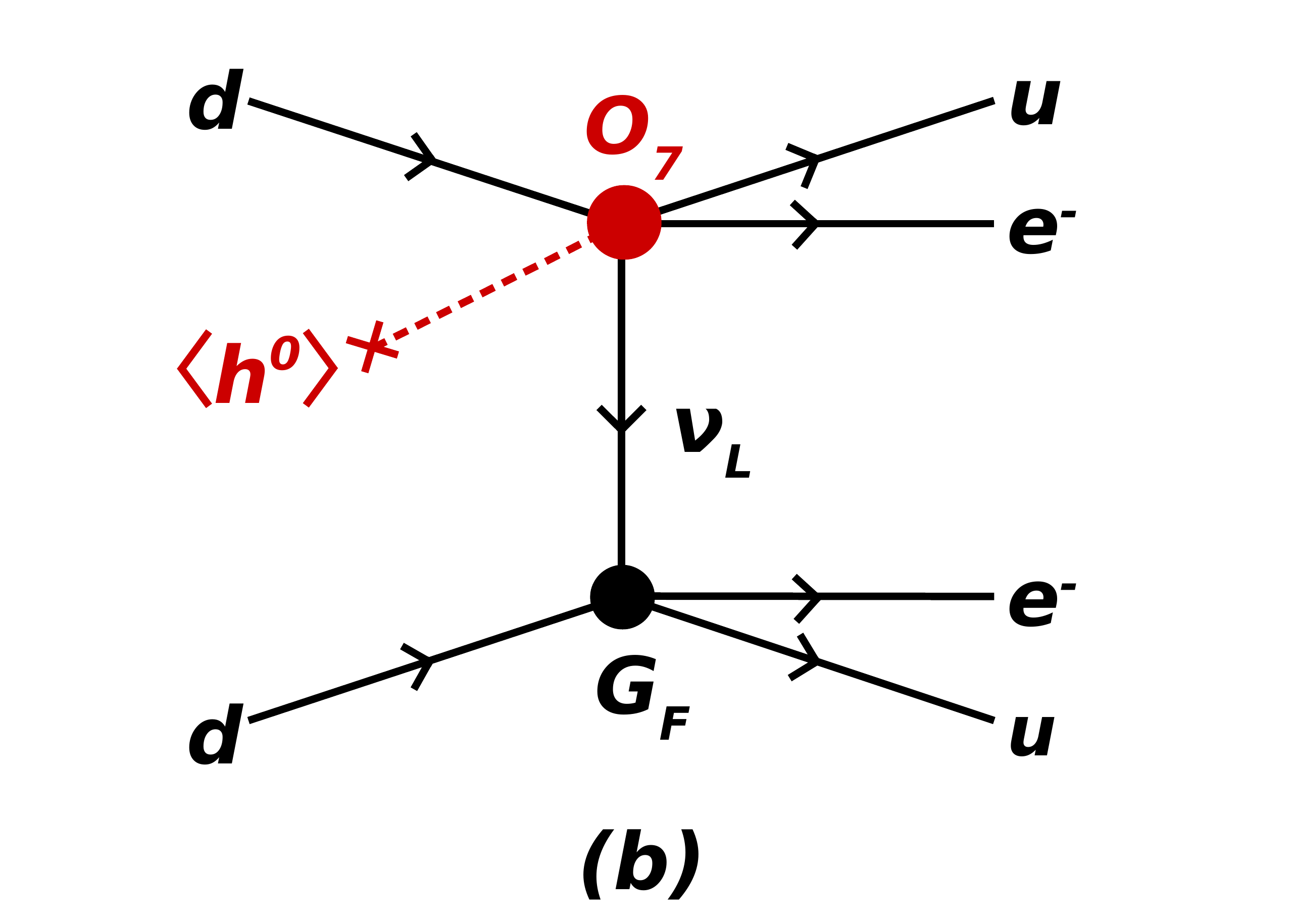}
	\includegraphics[clip,width=0.32\textwidth]{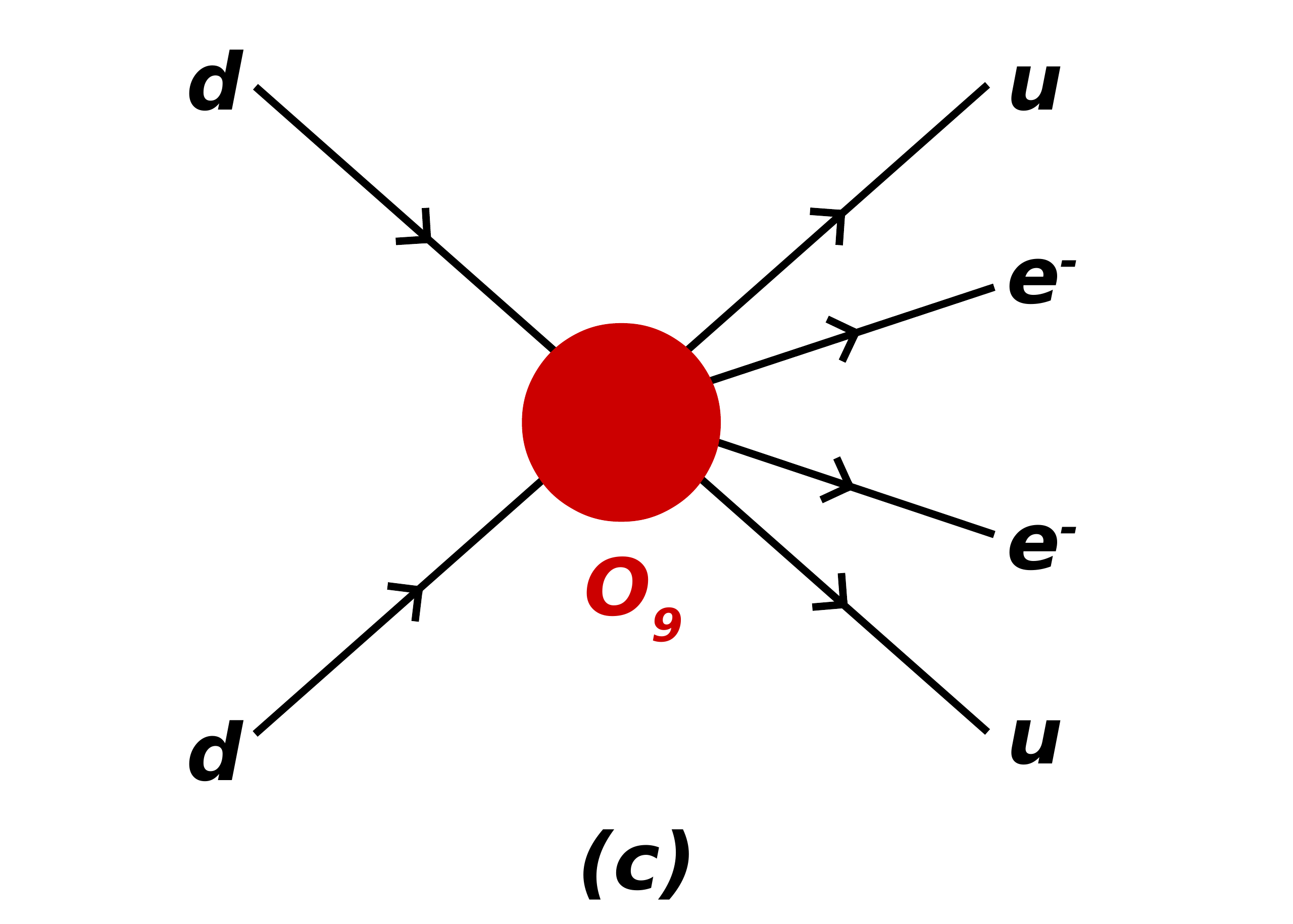}
	\caption{Contributions to $\ovbb$ decay from effective higher-dimensional LNV operators: (a) 5-dim Weinberg operator (standard mass mechanism), (b) 7-dim operator leading to long--range contribution, (c) 9-dim operator leading to short--range contribution. Adapted from \cite{Deppisch:2017ecm}.}
	\label{fig:contributions} 
\end{figure}
As a basis of our subsequent discussion, we provide a brief overview of the possible effective contact interactions at the Fermi scale $m_F \approx 100$~MeV at which $\ovbb$ decay occurs. These are likewise triggered by effective SM invariant operators violating $\Delta L = 2$ of dimension 5, 7, 9, 11, etc.. Fig.~\ref{fig:contributions} schematically shows the contribution of such operators. They can in general be categorized in two main classes: (i) Long-range transitions proceeding through the exchange of a light neutrino. This includes the so-called standard neutrino mass mechanism via Majorana neutrinos in Fig.~\ref{fig:contributions}~(a) but also via exotic interactions incorporating right-chiral neutrinos Fig.~\ref{fig:contributions}~(b). In the SM with only left-handed neutrinos, these operators violate $\Delta L = 2$ and they incorporate a helicity flip through inclusion of a Higgs field. (ii) Short-range transitions with no mediating particle lighter than $\approx 100$~MeV. As contact interactions with six external fermions, they are of dimension 9 and higher odd dimensions. The most prominent scenario where such an operator is generated is through the inclusion of heavy sterile neutrinos \cite{Simkovic:1999re}. In the above classification we do not include the case where additional light states are either mediating the decay or are emitted in it (e.g. Majorons).

Probing exotic $\Delta L = 2$ transitions is crucial for our understanding of how light neutrinos acquire their tiny masses. If exotic contributions were to be observed in upcoming experiments, it would indicate that the origin of light neutrino masses is around the TeV scale. Likewise, the non-observation of $\ovbb$ puts strong constraints on neutrino mass mechanisms close to the EW scale. It is not just neutrino physics that can be probed, though. Operators violating $\Delta L = 2$, or the underlying physics responsible for them, can also erase an asymmetry between the number of leptons and anti-leptons throughout the thermal history of the early universe. Together with the sphaleron transitions in the SM violating the sum of total baryon and lepton number $(B+L)$, this will also erase an asymmetry between baryons and anti-baryons. The rate of this washout can be related to the half life of $\ovbb$ decay for a given operator. The observation of non-standard $\ovbb$ decay mechanisms can thus generally falsify baryogenesis mechanisms operating at scales above the EW scale \cite{Deppisch:2015yqa, Deppisch:2017ecm}. A similar argument applies to other process probing lepton number violation around the TeV scale, such as searches for same sign di-leptons at the LHC \cite{Deppisch:2013jxa, Deppisch:2014hva}. 

Before discussing the exotic short-range contributions of our interest, we remind the reader that the mass mechanism of $0\nu\beta\beta$ decay is sensitive to the effective Majorana neutrino mass
\begin{align}
	m_\nu = \sum_{i=1}^3 U_{ei}^2 m_{\nu_i} \equiv m_{ee},
\end{align}
where the sum is over all light neutrinos with masses $m_{\nu_i}$, weighted by the square of the charged-current leptonic mixing matrix $U$. This quantity is equal to the $(ee)$ entry of the Majorana neutrino mass matrix. The inverse $0\nu\beta\beta$ decay half life in a given isotope can then be expressed by
\begin{align}
\label{eq:halflife}
	T_{1/2}^{-1} = \left|\frac{m_\nu}{m_e}\right|^2 G_\nu |\nme_\nu|^2,
\end{align}
where $G_\nu$ is the phase space factor (PSF) and $\nme_\nu$ the corresponding nuclear matrix element (NME) of the process. The normalization with respect to the electron mass $m_e$ yields a small dimensionless parameter $\epsilon_\nu = m_\nu/m_e$. The current experimental results lead to a limit $m_\nu \lesssim 0.06 - 0.17$~eV \cite{KamLAND-Zen:2016pfg}, with an uncertainty due to the different NMEs in various nuclear structure models. Future experiments will probe $m_\nu \approx 0.02$~eV, corresponding to the lowest value for an inverse hierarchy of the light neutrino states. A popular modification is through the inclusion of light sterile neutrinos with masses in the range from eV to MeV \cite{Pontecorvo:1968wp, Giunti:2010wz, Giunti:2013aea, Barry:2011wb}, in which case the half life is still given by Eq.~\eqref{eq:halflife}, but with different masses $m_{\nu_i}$ and couplings $U_{ei}$ \cite{Barea:2015zfa, Blennow:2010th, Faessler:2014kka, Deppisch:2004kn}.

The NMEs of the nuclear $\ovbb$ transitions are notoriously difficult to calculate and limits derived from $\ovbb$ decay are affected for any contribution. Detailed treatments using different nuclear structure model approaches can for example be found in \cite{Barea:2009zza, Barea:2013bz, Barea:2015kwa, Simkovic:2007vu, Simkovic:2013qiy, Suhonen:1991sk, Suhonen:2012wd, Caurier:2007xz, Menendez:2008jp, Rodriguez:2010mn}. Despite tremendous efforts to improve the nuclear theory calculation, the latest matrix elements obtained using various approaches differ in many cases by factors of ${\sim}(2-3)$. Experimentally, the most stringent bounds on $\ovbb$ decay are currently from $^{76}$Ge \cite{Agostini:2017iyd} and $^{136}$Xe \cite{KamLAND-Zen:2016pfg}. The results presented below are using the recent results in $^{76}$Ge of $T^\text{Ge}_{1/2} \ge 5.3\times 10^{25}$~y and in $^{136}$Xe of $T^\text{Xe}_{1/2} \ge 1.07\times 10^{26}$~y at $90\%$ confidence level (CL). Planned future experiments searching for $\ovbb$ decay are expected to reach sensitivities of the order of $T_{1/2} \approx 10^{27}$~y. For example, the recent comparative analysis \cite{Agostini:2017jim} quotes a discovery sensitivity at $3\sigma$ of $T^\text{Xe}_{1/2} = 4.1\times 10^{27}$~y for the planned nEXO experiment \cite{Mong:2016sza}. For more details on the effective $\ovbb$ interaction, see for example the review \cite{Deppisch:2012nb} and references therein. General up-to-date reviews of $\ovbb$ decay and associated physics can be found in \cite{Vergados:2016hso}, while a more specific recent review on $0\nu\beta\beta$ NMEs is available in \cite{Engel:2016xgb}.

Besides the light and heavy neutrino exchange, exotic long-range mechanisms have received most attention so far \cite{Doi:1981, Doi:1983, Tomoda:1990rs, Ali:2006iu, Ali:2007ec}. This is reasonable as the underlying SM operators already occur at dimension 7. It is important to note though that due to the helicity-flip involved, such operators are typically suppressed by the smallness of the light neutrino masses \cite{Helo:2016vsi}. In our work, we instead focus on short-range mechanisms that often do not suffer such a strong suppression at a similar level. In both cases, the half life triggered by a single mechanism may be generically expressed similarly to Eq.~\eqref{eq:halflife},
\begin{align}
	T_{1/2}^{-1} = |\epsilon_I|^2 G_I \abs{\nme_I}^2,
\end{align}
where $G_I$ is the nuclear PSF and $\nme_I$ the NME, both generally depending on the Lorentz structure of the effective operator in question. The coupling constant $\epsilon_I$ parametrizes the underlying particle physics dynamics, e.g. the couplings to and the masses of the heavy states integrated out. NMEs and PSFs for dimension-6 operators were given in \cite{Ali:2007ec}. In this paper we present a detailed derivation of NMEs and PSFs for dimension-9 effective operators.

The paper is arranged as follows. After presenting the general effective Lagrangian at the quark level in Section~\ref{sec:effLagrangian}, we outline the calculation of the $\ovbb$ differential decay rate in Section~\ref{sec:decayrate}. We then give the derivation of the NMEs in Section~\ref{sec:nme}. Section~\ref{sec:PSFs} details the calculation of the leptonic PSFs. The results of these calculations are then combined in Section~\ref{sec:numresults} to give explicit expressions for the decay rate and angular correlations. Limits on the effective couplings $\epsilon_I$ are also derived therein, assuming one contribution is different from zero at a time. Section~\ref{sec:summary} contains some concluding remarks.

\section{Effective Particle Physics Lagrangian}
\label{sec:effLagrangian}

The contributions to $0\nu\beta\beta$ decay can be parametrized by effective operators of dimension 6 and 9~\cite{Pas:1999fc, Pas:2000vn}, corresponding to short-range and long-range interactions, respectively. The general Lagrangian of $0\nu\beta\beta$ decay consists of long-range and short-range parts, corresponding to point-like vertices at the Fermi scale $\approx 100$~MeV.

In this work, we concentrate on the short-range contributions for which the general effective interaction Lagrangian schematically reads \cite{Pas:2000vn} 
\begin{align}
\label{eq:lagsr}
	\mathcal{L}_\text{SR} = 
	\frac{G^2_F}{2m_p}\sum_{\text{chiralities}}\sqb{
		  \epsilon_1^{\bullet} J_{\circ}J_{\circ} j_{\circ}
		+ \epsilon_2^{\bullet} J_{\circ}^{\mu\nu} J_{\circ\mu\nu} j_{\circ} 
		+ \epsilon_3^{\bullet} J_{\circ}^\mu J_{\circ\mu} j_{\circ} 
		+ \epsilon_4^{\bullet} J_{\circ}^\mu J_{\circ\mu\nu} j^\nu 
		+ \epsilon_5^{\bullet} J_{\circ}^\mu J_{\circ} j_\mu
	},
\end{align}
where the sum and the place holders $\circ$ indicate that the currents involved can have different chiralities and there is a separate effective coupling $\epsilon_i^{\bullet}$ for each such combination. Specifically, the hadronic and leptonic currents in Eq.~\eqref{eq:lagsr} are
\begin{gather}
\label{eq:allcurrents}
	J_{R/L}          = \bar u(1\pm\gamma_5) d, \quad
	J^\mu_{R/L}      = \bar u\gamma^\mu(1\pm\gamma_5) d, \quad
	J^{\mu\nu}_{R/L} = \bar u\sigma_{\mu\nu}(1\pm\gamma_5) d, \\
	j_{R/L}          = \bar e(1\pm\gamma_5) e^c, \quad
	j^\mu            = \bar e\gamma^\mu\gamma_5 e^c,
\end{gather}
with $\sigma_{\mu\nu} = \frac{i}{2} \left[\gamma_\mu,\gamma_\nu\right]$. The fields $u$, $d$ and $e$ are 4-component Dirac spinor operators representing the up-quark, down-quark and electron, respectively. The field $e^c = C e$ denotes the charge conjugate, corresponding to the fact that all lepton currents violate the electron lepton number by two units. While the currents involved in Eq.~\eqref{eq:lagsr} can have different chiralities as denoted in Eq.~\eqref{eq:allcurrents}, the results will not depend on many of the specific choices. 

As convention, the Lagrangian Eq.~\eqref{eq:lagsr} is normalized by the factor $G_F^2/(2m_p)$ with the Fermi constant $G_F$ and the proton mass $m_p$. As a result, the effective coupling constants $\epsilon_i$ are dimensionless.

In the Lagrangian Eq.~\eqref{eq:lagsr} one does not have to consider all possible combinations of the chiralities of the currents, as some of them are redundant or vanish. In order to prevent any confusion about the basis of low-energy dimension-9 operators we are considering, we spell these explicitly out in Table \ref{tab:opbasis}. Each operator is labelled in the same way as the corresponding effective coupling in Eq.~\eqref{eq:lagsr}, i.e., $\mathcal{O}^{\bullet}_{i} \sim \epsilon^{\bullet}_{i}$, where the superscript specifies the chiralities of the particular bilinears in their respective order. We explicitly identify the equivalent (and thus redundant) operators and we omit operators $\mathcal{O}_2^{RLL}$, $\mathcal{O}_2^{LRL}$, $\mathcal{O}_2^{RLR}$ and $\mathcal{O}_2^{LRR}$, which trivially vanish, because of the identity
\begin{align}
	\sqb{u\gamma^\mu(1+\gamma_5)d} \sqb{u\gamma_\mu(1-\gamma_5)d} \equiv \sqb{u\gamma^\mu(1-\gamma_5)d} \sqb{u\gamma_\mu(1+\gamma_5)d} = 0.
\end{align}
Similarly, the Lagrangian Eq.~\eqref{eq:lagsr} does not contain any terms with vector, tensor or axial-tensor electron currents, as $\bar e\gamma^\mu e^c = 0$ and $\bar e\sigma_{\mu\nu}(1\pm\gamma_5) e^c = 0$, due to the Pauli exclusion principle.

\begin{table}
	\setlength{\tabcolsep}{0.25em}
	\renewcommand{\arraystretch}{1.1}
	\begin{tabular}{l | c} \hline
		$\mathcal{O}_1^{RRR}$ & $\sqb{\bar u^i(1+\gamma_5)d_i} \sqb{\bar u^j(1+\gamma_5)d_j} \sqb{\bar e(1+\gamma_5)e^c}$ \\
		$\mathcal{O}_1^{RRL}$ & $\sqb{\bar u^i(1+\gamma_5)d_i} \sqb{\bar u^j(1+\gamma_5)d_j} \sqb{\bar e(1-\gamma_5)e^c}$ \\
		$\mathcal{O}_1^{LRR} \equiv \mathcal{O}_1^{RLR}$ & $\sqb{\bar u^i(1-\gamma_5)d_i} \sqb{\bar u^j(1+\gamma_5)d_j} \sqb{\bar e(1+\gamma_5)e^c}$ \\
		$\mathcal{O}_1^{LRL} \equiv \mathcal{O}_1^{RLL}$ & $\sqb{\bar u^i(1-\gamma_5)d_i} \sqb{\bar u^j(1+\gamma_5)d_j} \sqb{\bar e(1-\gamma_5)e^c}$ \\
		$\mathcal{O}_1^{LLR}$ & $\sqb{\bar u^i(1-\gamma_5)d_i} \sqb{\bar u^j(1-\gamma_5)d_j} \sqb{\bar e(1+\gamma_5)e^c}$ \\
		$\mathcal{O}_1^{LLL}$ & $\sqb{\bar u^i(1-\gamma_5)d_i} \sqb{\bar u^j(1-\gamma_5)d_j} \sqb{\bar e(1-\gamma_5)e^c}$ \\ \hline
		
		$\mathcal{O}_2^{RRR}$ & $\sqb{\bar u^i\sigma^{\mu\nu}(1+\gamma_5)d_i} \sqb{\bar u^j\sigma_{\mu\nu}(1+\gamma_5)d_j} \sqb{\bar e(1+\gamma_5)e^c}$ \\
		$\mathcal{O}_2^{RRL}$ & $\sqb{\bar u^i\sigma^{\mu\nu}(1+\gamma_5)d_i} \sqb{\bar u^j\sigma_{\mu\nu}(1+\gamma_5)d_j} \sqb{\bar e(1-\gamma_5)e^c}$ \\
		$\mathcal{O}_2^{LLR}$ & $\sqb{\bar u^i\sigma^{\mu\nu}(1-\gamma_5)d_i} \sqb{\bar u^j\sigma_{\mu\nu}(1-\gamma_5)d_j} \sqb{\bar e(1+\gamma_5)e^c}$ \\
		$\mathcal{O}_2^{LLL}$ & $\sqb{\bar u^i\sigma^{\mu\nu}(1-\gamma_5)d_i} \sqb{\bar u^j\sigma_{\mu\nu}(1-\gamma_5)d_j} \sqb{\bar e(1-\gamma_5)e^c}$ \\ \hline
		
		$\mathcal{O}_3^{RRR}$ & $\sqb{\bar u^i\gamma^{\mu}(1+\gamma_5)d_i} \sqb{\bar u^j\gamma_{\mu}(1+\gamma_5)d_j} \sqb{\bar e(1+\gamma_5)e^c}$ \\
		$\mathcal{O}_3^{RRL}$ & $\sqb{\bar u^i\gamma^{\mu}(1+\gamma_5)d_i} \sqb{\bar u^j\gamma_{\mu}(1+\gamma_5)d_j} \sqb{\bar e(1-\gamma_5)e^c}$ \\
		$\mathcal{O}_3^{LRR} \equiv \mathcal{O}_3^{RLR}$ & $\sqb{\bar u^i\gamma^{\mu}(1-\gamma_5)d_i} \sqb{\bar u^j\gamma_{\mu}(1+\gamma_5)d_j} \sqb{\bar e(1+\gamma_5)e^c}$ \\
		$\mathcal{O}_3^{LRL} \equiv \mathcal{O}_3^{RLL}$ & $\sqb{\bar u^i\gamma^{\mu}(1-\gamma_5)d_i} \sqb{\bar u^j\gamma_{\mu}(1+\gamma_5)d_j} \sqb{\bar e(1-\gamma_5)e^c}$ \\
		$\mathcal{O}_3^{LLR}$ & $\sqb{\bar u^i\gamma^{\mu}(1-\gamma_5)d_i} \sqb{\bar u^j\gamma_{\mu}(1-\gamma_5)d_j} \sqb{\bar e(1+\gamma_5)e^c}$ \\
		$\mathcal{O}_3^{LLL}$ & $\sqb{\bar u^i\gamma^{\mu}(1-\gamma_5)d_i} \sqb{\bar u^j\gamma_{\mu}(1-\gamma_5)d_j} \sqb{\bar e(1-\gamma_5)e^c}$ \\ \hline
		
		$\mathcal{O}_4^{RR}$ & $\sqb{\bar u^i\gamma^{\mu}(1+\gamma_5)d_i} \sqb{\bar u^j\sigma_{\mu\nu}(1+\gamma_5)d_j} \sqb{\bar e\gamma^{\nu}\gamma_5 e^c}$ \\
		$\mathcal{O}_4^{RL}$ & $\sqb{\bar u^i\gamma^{\mu}(1+\gamma_5)d_i} \sqb{\bar u^j\sigma_{\mu\nu}(1-\gamma_5)d_j} \sqb{\bar e\gamma^{\nu}\gamma_5 e^c}$ \\
		$\mathcal{O}_4^{LR}$ & $\sqb{\bar u^i\gamma^{\mu}(1-\gamma_5)d_i} \sqb{\bar u^j\sigma_{\mu\nu}(1+\gamma_5)d_j} \sqb{\bar e\gamma^{\nu}\gamma_5 e^c}$ \\
		$\mathcal{O}_4^{LL}$ & $\sqb{\bar u^i\gamma^{\mu}(1-\gamma_5)d_i} \sqb{\bar u^j\sigma_{\mu\nu}(1-\gamma_5)d_j} \sqb{\bar e\gamma^{\nu}\gamma_5 e^c}$ \\ \hline
		
		$\mathcal{O}_5^{RR}$ & $\sqb{\bar u^i\gamma^{\mu}(1+\gamma_5)d_i} \sqb{\bar u^j(1+\gamma_5)d_j} \sqb{\bar e\gamma_{\mu}\gamma_5 e^c}$ \\
		$\mathcal{O}_5^{RL}$ & $\sqb{\bar u^i\gamma^{\mu}(1+\gamma_5)d_i} \sqb{\bar u^j(1-\gamma_5)d_j} \sqb{\bar e\gamma_{\mu}\gamma_5 e^c}$ \\
		$\mathcal{O}_5^{LR}$ & $\sqb{\bar u^i\gamma^{\mu}(1-\gamma_5)d_i} \sqb{\bar u^j(1+\gamma_5)d_j} \sqb{\bar e\gamma_{\mu}\gamma_5 e^c}$ \\
		$\mathcal{O}_5^{LL}$ & $\sqb{\bar u^i\gamma^{\mu}(1-\gamma_5)d_i} \sqb{\bar u^j(1-\gamma_5)d_j} \sqb{\bar e\gamma_{\mu}\gamma_5 e^c}$ \\ \hline
	\end{tabular}
	\caption{Basis of low-scale, 9-dimensional operators invariant under the $SU(3)_C \otimes U(1)_{Q}$ gauge group contributing to $\ovbb$ decay.}
	\label{tab:opbasis}
\end{table}
The 24 operators in Tab.~\ref{tab:opbasis} are linearly independent and form a complete basis of 9-dimensional operators invariant under the gauge group $SU(3)_C \otimes U(1)_{Q}$ and contributing to $\ovbb$ decay. We also show explicitly the assumed contractions of the colour indices $i$, $j$, although these are trivial, as always the quarks within the same Lorentz bilinear are contracted. The total number of these operators agrees with the result we obtained as a consistency check from a calculation using the Hilbert series method \cite{Lehman:2014jma, Henning:2015alf} and with the results in \cite{Graesser:2016bpz}. Despite comments in the latter reference we do not see the need to include operators containing quark bilinears transforming as colour octets. These can be shown to be related by Fierz transformation to the herein presented operators with tensor Lorentz structure. For example, for operator $\mathcal{O}_2^{LLL}$ one can find the following Fierz identity
\begin{align} 
\label{eq:op2fierz}
	\mathcal{O}_2^{LLL} &= 
	\sqb{\bar u^i\sigma^{\mu\nu}(1-\gamma_5)d_i} 
	\sqb{\bar u^j\sigma_{\mu\nu}(1-\gamma_5)d_j} j_L \nonumber\\
	&= 2 \sqb{\bar u^i(1-\gamma_5)d_j}\sqb{\bar u^j(1-\gamma_5)d_i} j_L
	- \sqb{\bar u^i(1-\gamma_5)d_i} \sqb{\bar u^j(1-\gamma_5)d_j} j_L.
\end{align}
If we further apply the $SU(3)_C$ colour Fierz identity based on the well-known group-theoretical formula
\begin{align}
	\delta_{ij}\delta_{kl} = 
	\frac{1}{3}\delta_{il}\delta_{kj} + \frac{1}{2}(\lambda^a)_{il}(\lambda_a)_{kj},
\end{align}
on the first term on the right-hand side of Eq.~\eqref{eq:op2fierz}, we get
\begin{align}
	\mathcal{O}_2^{LLL}
	&= \sqb{\bar u^i(1-\gamma_5)(\lambda^a)_{ik}d^k}
	\sqb{\bar u^j(1-\gamma_5)(\lambda_a)_{jl}d^l} j_L 
	- \frac{1}{3} \sqb{\bar u^i(1-\gamma_5)d_i}
	\sqb{\bar u^j(1-\gamma_5)d_j} j_L \nonumber\\
	&\equiv \sqb{\bar u^i(1-\gamma_5)(\lambda^a)_{ik}d^k}
	\sqb{\bar u^j(1-\gamma_5)(\lambda_a)_{jl}d^l} j_L 
	- \frac{1}{3} \mathcal{O}^{LLL}_1.
\end{align}
Hence, we see that if the operator containing colour octets (the first term on the right-hand side of the above equation) is neglected, for the fact it does not contribute to $\ovbb$ decay as argued in \cite{Prezeau:2003xn}, then the operator $\mathcal{O}_2^{LLL}$ is equivalent to $\mathcal{O}_1^{LLL}$, and thus redundant. In a similar way all the operators with tensor quark bilinears can be traded for operators containing colour octets. If these are neglected, then only the operators consisting of (both left- and right-handed) scalar and vector colour-singlet quark bilinears are left.

\section{Neutrinoless Double Beta Decay Rate}
\label{sec:decayrate}

The differential rate of $\ovbb$ decay can be written as 
\begin{align}
	d\Gamma = 2\pi \overline{|\mathcal{R}|^2} \delta(E_1 + E_2 + E_F - E_I) 
	     \frac{d^3\vecl{p}_1}{(2\pi)^3}\frac{d^3\vecl{p}_2}{(2\pi)^3},
\end{align}
where $\overline{|\mathcal{R}|^2}$ is the full matrix element of the $\ovbb$ decay process summed over the spin projections $s_1$, $s_2$ of the electrons and the final nuclear state $S_F$. The 4-momenta of the outgoing electrons are $(E_1,\vecl{p}_1)$, $(E_2,\vecl{p}_2)$, and $E_F$ and $E_I$ are the energies of the final and initial nuclei, respectively. The $Q_{\beta\beta}$ value of the transition, i.e. the kinetic energy release of the electrons, is given by $Q_{\beta\beta} = E_I - E_F - 2 m_e$, with the electron mass $m_e = 0.511$~MeV. Here, we neglect the recoil energy of the final nucleus which is of the order $Q_{\beta\beta}^2/(2M_A) = \mathcal{O}(0.1~\text{keV})$ for isotope masses $M_A$ of interest. Because of the overall rotational invariance and energy conservation, the differential rate can be expressed in terms of the energy $m_e < E_1 < Q_{\beta\beta} + m_e$ of one of the electrons and the angle $0 \leq \theta \leq \pi$ between the two electrons, with $\cos\theta = \vecl{\hat p}_1 \cdot \vecl{\hat p}_2$.\footnote{Throughout, we denote normalized vectors by $\vecl{\hat v} \equiv \vecl{v} / |\vecl{v}|$.} The energy of the other electron is then determined as $E_2 = Q_{\beta\beta} + 2m_e - E_1$ and the magnitudes of the electron 3-momenta are $p_i \equiv |\vecl{p}_i| = \sqrt{E_i^2 - m_e^2}$. 

The full matrix element of the process can be formally expressed as
\begin{align}
\label{eq:fullnme}
	\mathcal{R} = 
	\bra{\mathcal{O}_F^+ e_{\vecl{p}_1 s_1} e_{\vecl{p}_2 s_2}} 
	\mathcal{L}_\text{SR} \ket{\mathcal{O}_I^+}.
\end{align}
Here, $\bra{\mathcal{O}_F^+ e_{\vecl{p}_1 s_1} e_{\vecl{p}_2 s_2}}$ denotes the final state composed of the $0^+$ daughter nuclear state and the two emitted electrons, and $\ket{\mathcal{O}_I^+}$ the initial nuclear state. It is understood that the wavefunction of the two electrons in Eq.~\eqref{eq:fullnme} is anti-symmetrized; the same holds for the wavefunctions of the $\mathcal{O}_I^+$ and $\mathcal{O}_F^+$ states in terms of their constituent nucleons. In Eq.~\eqref{eq:fullnme}, we allow the most general form of the quark level Lagrangian Eq.~\eqref{eq:lagsr} for $\mathcal{L}_\text{SR}$, which we symbolically express as
\begin{align}
	\mathcal{L}_\text{SR} = \sum_{K,\Xi} \epsilon_K j_K^\Xi J_K^\Xi J_K^{'\Xi},
\end{align}
where the summation over $K$ and $\Xi$ collectively denotes the different electron-quark-quark current combinations $jJJ'$ (including different chiralities) and the Lorentz contractions, respectively.

The evaluation of the matrix element $\mathcal{R}$ is rather complicated since, in general, the leptonic part is nested with the hadronic part. For the long-range case a detailed calculation was given by Doi et al. \cite{Doi:1981, Doi:1983} and Tomoda \cite{Tomoda:1987ji}. Since the hadronic part is the product of two currents, a sum over a set of intermediate states $\ket{\mathcal{N}}$ must be performed. This is a daunting task since for $\ovbb$ decay all states up to an energy $E \approx 100$~MeV contribute. It is therefore customary to treat the above summation in the \emph{closure approximation}, i.e. sum over a complete set of states,
\begin{align}
	\sum_\mathcal{N} \bra{\mathcal{O}_F^+} J_K^\Xi \ket{\mathcal{N}}\bra{\mathcal{N}} J_K^{'\Xi} \ket{\mathcal{O}_I^+} \approx \bra{\mathcal{O}_F^+} J_K^\Xi J_K^{'\Xi} \ket{\mathcal{O}_I^+}.
\end{align}
This approximation is very well justified in the case of short-range operators as the intermediate transition occurs at very high energies $|\vecl{q}| \approx 100$~MeV, corresponding to the inter-nucleon distance, compared to the nuclear transition itself at $Q_{\beta\beta} \approx 1$~MeV.

Another problem is that the leptonic and hadronic parts are entangled. In order to disentangle them, an approximation is made wherein the electron wave functions are evaluated at the surface of the nucleus \cite{Doi:1981, Kotila:2012zza}. This approximation can be improved by using simplified nucleon wave functions and calculating the weighted average electron position \cite{Kotila:2012zza}, but the approximation we employ here does not introduce a sizeable error. 

Following this approach, the overall matrix element $\mathcal{R}$ is factorized into (i) the product of the leptonic matrix element that will be integrated over the two-electron phase space yielding the so-called phase space factor (PSF) that depends only on the leptonic current and the electron wave function at the surface of the nucleus\footnote{In the associated $2\nu\beta\beta$ decay it will also depend on the outgoing neutrino wave functions.}, and (ii) the nuclear matrix element (NME). 

For the latter, we first reduce each nucleon current $J_K^\Xi$ to its non-relativistic form by means of a Foldy-Wouthuysen (FW) transformation, then take the product of the two currents and evaluate the matrix elements of the corresponding two-body operator in the nuclear many-body wavefunctions. In the FW transformation we take terms to order $|\vecl{q}|/m_p$. This is a good approximation, since the momentum transfer in the process is of order $|\vecl{q}| \sim 100$~MeV, and therefore $|\vecl{q}|/m_p \sim 0.1$. In certain cases of enhanced form factors higher order terms in the products of hadronic currents are also taken into account.

Altogether, this yields the full matrix element
\begin{align}
\label{eq:fillmatrixelement}
	\mathcal{R} = \frac{G_F^2}{2m_p} 
	    \sum_{K,\Xi} \epsilon_K 
	    \bra{e_{\vecl{p}_1s_1}}
	    	j_K^\Xi \ket{e^c_{\vecl{p}_2s_2}}
	    \bra{\mathcal{O}_F^+} J_K^\Xi J_K^{'\Xi} \ket{\mathcal{O}_I^+}.
\end{align}
The leptonic matrix elements will be evaluated using the appropriate electron wavefunctions in Section~\ref{sec:PSFs}. The nucleon matrix elements are evaluated as discussed above, including appropriate $\vecl{q}^2$-dependent form factors, in Section~\ref{sec:quarktonucleons}-B, and the nuclear matrix elements in Section~\ref{sec:nucleonstonucleus}.

Putting together the PSFs and NMEs, one can write the fully differential rate for $0^+\to 0^+$ $\ovbb$ decay as \cite{Doi:1981, Doi:1983, Tomoda:1990rs}\footnote{We note in passing that for $0^+\to 2^+$ $\ovbb$ decay, there is an additional term in Eq.~\eqref{eq:differentialrate} of the form $c(E_1)(\cos^2\theta - 1/3)$.}
\begin{align}
\label{eq:differentialrate}
	\frac{d^2\Gamma}{dE_1 d\cos\theta} = 
	C\, w(E_1) \left(a(E_1) + b(E_1)\cos\theta\right),
\end{align}
with
\begin{align}
	C      &= \frac{G_F^4 m_e^2}{16\pi^5}, \label{eq:C} \\
	w(E_1) &= E_1 E_2 p_1 p_2, \label{eq:w}
\end{align}
where $E_2$, $p_1$ and $p_2$ are expressed as functions of $E_1$.

Following the notation of \cite{Doi:1981, Doi:1983, Tomoda:1990rs} the coefficient $1/m_p$ appearing in Eq.~\eqref{eq:fillmatrixelement} is included in the calculation of the nuclear matrix elements, see the following Eq.~\eqref{eq:neutrinopotential} and a mass $m_e$ is added in the numerator to cancel the mass $m_e$ in the denominator of the so-called neutrino potential in Eq.~\eqref{eq:neutrinopotential}.

The total decay rate $\Gamma$ and the decay half life $T_{1/2}$ are then given by
\begin{align}
\label{eq:totalrate}
	\Gamma = \frac{\ln 2}{T_{1/2}} = 2 C\int_{m_e}^{Q_{\beta\beta}+m_e} dE_1 w(E_1) a(E_1).
\end{align}
From Eq.~\eqref{eq:differentialrate}, one can calculate the single electron energy distribution,
\begin{align} \label{eq:energydistro}
	\frac{d\Gamma}{dE_1} = 2C w(E_1)a(E_1),
\end{align}
and the energy-dependent angular correlation
\begin{align}
	\alpha(E_1) = \frac{b(E_1)}{a(E_1)}.
\end{align}
Introducing the integrated quantities
\begin{align} 
	A = \int_{m_e}^{Q_{\beta\beta}+m_e} dE_1 w(E_1) a(E_1), \quad
	B = \int_{m_e}^{Q_{\beta\beta}+m_e} dE_1 w(E_1) b(E_1),
\end{align}
and their ratio $K = B/A$, one obtains the angular distribution 
\begin{align}
	\frac{d\Gamma}{d\cos\theta} = \frac{\Gamma}{2}\left(1 + K\cos\theta\right).
\end{align}
Although both the single electron energy distribution and the angular correlation require dedicated experimental setups, we calculate them nonetheless since they contain important information on the underlying mechanism.

\section{Nuclear Matrix Elements}
\label{sec:nme}

\subsection{From Quarks to Nucleons}
\label{sec:quarktonucleons}
We are interested in matrix elements induced by the quark bilinears appearing in the Lagrangian Eq.~\eqref{eq:lagsr}, i.e. by the left- and right-handed scalar, vector and tensor quark currents. Considering the nucleon isodoublet $N = \binom{p}{n}$, the nucleon matrix elements of these colour singlet quark currents have according to \cite{Adler:1975he} the following structure
\begin{align}
\label{eq:nucleoncurrents}
	\bra{p}\bar{u}(1\pm\gamma_5)d\ket{n} &= 
	\bar{N} \tau^+ 
	\sqb{F_S(q^2) \pm F_{PS}(q^2)\gamma_5} N', \\
	\bra{p}\bar{u}\gamma^\mu(1\pm\gamma_5)d\ket{n} &= 
	\bar{N} \tau^+ 
	\sqb{F_V(q^2)\gamma^\mu-i\frac{F_W(q^2)}{2m_p}\sigma^{\mu\nu}q_\nu}N' 
	\nonumber\\ \label{eq:vecNucleons}
	&\pm \bar{N} \tau^+ \sqb{F_A(q^2)\gamma^\mu\gamma_5 
	- \frac{F_P(q^2)}{2m_p}\gamma_5q^\mu}N', \\
	\bra{p}\bar{u}\sigma^{\mu\nu}(1\pm\gamma_5)d\ket{n} &= 
	\bar{N} \tau^+ \sqb{J^{\mu\nu} \pm \frac{i}{2} \epsilon^{\mu\nu\rho\sigma} J_{\rho\sigma}} N'.
	\label{eq:tenNucleons}
\end{align}
Here we define
\begin{align}
	J^{\mu\nu} = F_{T_1}(q^2)\sigma^{\mu\nu} + i\frac{F_{T_2}(q^2)}{m_p}(\gamma^\mu q^\nu 
	- \gamma^\nu q^\mu) + \frac{F_{T_3}(q^2)}{m_p^2}(\sigma^{\mu\rho} q_\rho q^\nu 
	- \sigma^{\nu\rho} q_\rho q^\mu),
\end{align}
and $\tau^+$ is the isospin-raising operator, transforming a neutron into a proton. The matrix elements are in general functions of the neutron and proton momenta $p_n = p_{N'}$ and $p_p = p_N$, respectively, and the momentum transfer entering the form factors is defined as $q = p_p - p_n$. In Eq.~\eqref{eq:vecNucleons} we omit the induced scalar and axial-tensor terms; the corresponding currents can be safely neglected because they vanish in the isospin-symmetric limit \cite{PhysRev.112.1375}. Moreover, they are suppressed by a factor $1/m_p$ and they are not enhanced by a pion resonance. 

Important ingredients in Eqs.~\eqref{eq:nucleoncurrents}-\eqref{eq:tenNucleons} are the $q^2$-dependent form factors $F_X(q^2)$ with $X \in \{S, PS, V, W, A, P, T_1, T_2, T_3\}$. We parametrize these except $F_{PS}(q^2)$ and $F_P(q^2)$ in the so-called dipole form,
\begin{align}
\label{eq:formfactor_dipole}
	F_X(q^2) = \frac{g_X}{\left(1 + q^2 / m_X^2\right)^2},
\end{align}
where the coupling constants $g_X$ give the value of the form factor at zero momentum transfer, $g_X = F_X(0)$.

For example, the vector form factor can be experimentally determined from the electromagnetic form factor and from the conserved vector current (CVC) hypothesis,
\begin{align}
\label{eq:formfactor_V}
	F_V(q^2) = \frac{g_V}{\left(1 + q^2 / m_V^2\right)^2}, 
	\quad g_V = 1, \quad m_V = 0.84~\text{GeV}.
\end{align}
This parametrization provides a good description of $F_V(q^2)$ in the range $0 \leq |q| \leq 200$~MeV of interest in $\ovbb$ decay. A better parametrization, important for large $q^2 \gtrsim 1$~GeV$^2$, is given in \cite{Bijker:2004yu}, but it is of no interest for the purposes of the present paper. 

The induced form factor $F_W(q^2)$ can also be determined from experiment, since it is related to the Pauli form factor $F_2(q^2)$ \cite{Bijker:2004yu} and to the isovector anomalous magnetic moment of the nucleon,
\begin{align}
\label{eq:formfactor_W}
	F_W(q^2) = \frac{g_W}{\left(1 + q^2 / m_W^2\right)^2}, 
	\quad g_W = \mu_p - \mu_n = 3.70, \quad m_W = m_V = 0.84~\text{GeV},
\end{align}
where $\mu_p - \mu_n$ is the anomalous isovector magnetic moment of the proton and neutron.

The axial vector form factor can also be parametrized in dipole form and it is obtained from experiment,
\begin{align}
\label{eq:formfactor_A}
	F_A(q^2) = \frac{g_A}{\left(1 + q^2 / m_A^2\right)^2}, 
	\quad g_A = 1.269, \quad m_A = 1.09~\text{GeV}.
\end{align}
The value of $g_A$ is determined in neutron decay \cite{Yao:2006px} and $m_A$ is obtained from neutrino scattering \cite{Schindler:2006jq}.

The induced form factor $F_P(q^2)$ cannot be directly obtained from experiment. We use the parametrization suggested in \cite{Simkovic:1999re}, based on the partially conserved axial-vector current (PCAC) hypothesis,
\begin{align}
\label{eq:formfactor_P}
	F_P(q^2) = \frac{g_A}{\left(1 + q^2 / m_A^2\right)^2}
	\frac{1}{1 + q^2 / m_\pi^2}\frac{4m_p^2}{m_\pi^2}
	\left(1-\frac{m_\pi^2}{m_A^2}\right),
\end{align}
with the pion mass $m_\pi = 0.138$~GeV.

From Eq.~\eqref{eq:formfactor_P} we have $g_P\equiv F_P(0)=231$. This formula is consistent with a recent analysis in chiral perturbation theory \cite{Bernard:2001rs}, which gives $g_P=233$ and with recent measurements in muon capture, which give $F_P(q^2)\frac{|\vecl{q}^2|}{2m_p}$ at $|\vecl{q}|=0.88m_{\mu}$, where $m_{\mu}=0.105$ GeV is the muon mass. The calculated value is $8.0$, while the measured value is $8.06\pm0.55$ \cite{Andreev:2012fj}.

A considerable amount of attention has been devoted recently to the form factors $F_S(q^2)$ and $F_{PS}(q^2)$, in particular to the values at zero momentum transfer, $g_S = F_S(0)$ and $g_{PS} = F_{PS}(0)$. Quoted values are $g_S = 1.02\pm 0.11$ and $g_{PS} = 349\pm 9$ \cite{Gonzalez-Alonso:2018omy}. Not much is known about the $q^2$-dependence; for the scalar form factor, which, in the Breit frame, is the Fourier transform of the matter distribution, a reasonable parametrization is in dipole form with $g_S = 1$ and $m_S = m_V = 0.84$~GeV,
\begin{align}
\label{eq:formfactor_S}
	F_S(q^2) = \frac{g_S}{\left(1 + q^2 / m_S^2\right)^2},\quad 
	g_S = 1,\quad m_S = m_V = 0.84~\text{GeV}.
\end{align}

The value of the pseudo-scalar form factor $F_{PS}(q^2)$ diverges at $q^2 = 0$ in the chiral limit and results of lattice calculations depend on the extrapolation procedure. We take 
\begin{align}
\label{eq:formfactor_PS}
	F_{PS}(q^2) = 
	\frac{g_{PS}}{\left(1 + q^2/m_{PS}^2\right)^2}
	\frac{1}{1 + q^2/m^2_\pi}, 
	\quad g_{PS} = 349, \quad  m_{PS} = m_V = 0.84~\text{GeV}.
\end{align}
The question of whether or not the value of $g_{PS}$ is enhanced as in Ref.~\cite{Gonzalez-Alonso:2018omy} is beyond the scope of this paper. The parametrization \eqref{eq:formfactor_PS} reduces to the simple monopole form $1/\left(1 + q^2/m^2_\pi\right)$ used in chiral perturbation theory, but it includes the finite size of the nucleon.

No experimental information is available for the tensor form factors. Ref.~\cite{Gonzalez-Alonso:2018omy} quotes a value of $0.987\pm 0.055$ for $F_{T_1}(0) \equiv g_{T_1}$. An old calculation \cite{Adler:1975he} estimates, from the MIT bag model, $F_{T_1}(0) \equiv g_{T_1} = 1.38$, $F_{T_2}(0) \equiv g_{T_2}=-3.30$ and $F_{T_3}(0) \equiv g_{T_3} = 1.34$. In this paper we take
\begin{align}
\label{eq:formfactor_T}
	F_{T_i}(q^2) = \frac{g_{T_i}}{\left(1 + q^2/m_{T_i}^2\right)^2}, 
	\quad  m_{T_i} = m_V = 0.84~\text{GeV}.
\end{align}
with $g_{T_1} = 1$ (and $g_{T_2} = -3.30$, $g_{T_3} = 1.34$ estimated from \cite{Adler:1975he}). The two form factors $F_{T_2}(q^2)$ and $F_{T_3}(q^2)$ do not enter the results of this paper, but are quoted here for completeness.

\subsection{Non-Relativistic Expansion}
\label{sec:nonrelapprox}
To obtain the nuclear matrix elements of interest, we have to calculate the non-relativistic expansion of the above nucleon matrix elements. This form is obtained by a Foldy-Wouthuysen transformation \cite{Foldy:1949wa, Rose:1954zza}, which is an expansion in powers of the velocity $v/c$ or equivalently in $|\vecl{p}|/m_p$.

The resulting expressions are summarized in the following section where we use the spatial momentum difference $\mathbf{q} = \mathbf{p}_p - \mathbf{p}_n$ and momentum sum $\mathbf{Q} = \mathbf{p}_p + \mathbf{p}_n$. 
Particular terms will be listed according to the order in $|\vecl{q}| / m_p$, where we perform the expansion up to and including terms of order $|\vecl{q}| / m_p$, except for terms incorporating $F_{P}(q^2)$ and $F_{PS}(q^2)$ which are enhanced as discussed above. In these cases we retain terms of order $\vecl{q}^2 / m_p^2$ and even higher.

\paragraph*{Scalar Bilinears} The non-relativistic expansion of the scalar and pseudo-scalar nucleon current corresponding to $J_{S\pm P} = (1\pm\gamma_5)$ can be written as
\begin{align} \label{eq:sbilinear}
	J_{S\pm P} = 
	F_S(q^2) I \pm \frac{F_{PS}(q^2)}{2m_p} 
	\boldsymbol{\sigma} \cdot \mathbf{q} + \dots.
\end{align}
Here, $I$ denotes the $2\times 2$ identity matrix and $\boldsymbol{\sigma} = (\sigma_1, \sigma_2, \sigma_2)^T$ is the vector of Pauli matrices, both operating in the spin space of the nucleon.

\paragraph*{Vector Bilinears}
The vector currents corresponding to $J^\mu_{V\pm A} = \gamma^\mu(1\pm\gamma_5)$ have four different components (vector, axial vector and induced pseudo-scalar and weak magnetism), which can be non-relativistically expanded as follows
\pagebreak
\begin{align} \label{eq:vbilinear}
	J^\mu_{V\pm A} &=
	g^{\mu i}\left[\mp F_A(q^2)\sigma_{i} 
	- \frac{F_V(q^2)}{2m_p}\mathbf{Q}_i I 
	+ \frac{F_V(q^2) + F_W(q^2)}{2m_p}i(\boldsymbol{\sigma} \times \mathbf{q})_i 
	\pm \frac{F_P(q^2)}{4m_p^2}q_i \boldsymbol{\sigma} \cdot \mathbf{q}\right] \\
	&+g^{\mu 0}\left[F_V(q^2) I 
	\pm \frac{F_A(q^2)}{2m_p} \boldsymbol{\sigma} \cdot \mathbf{Q} 
	\mp \frac{F_P(q^2)}{4m_p^2} q_0 \boldsymbol{\sigma} \cdot \mathbf{q} 
	\right] + \dots. \nn
\end{align}

\paragraph*{Tensor Bilinears} The non-zero nuclear components corresponding to the tensor bilinears $J^{\mu\nu}_{T\pm T_5} = \sigma^{\mu\nu}(1\pm\gamma_5)$ are
\begin{align} \label{eq:tbilinear}
	J^{\mu\nu}_{T\pm T_{5}} &=
	F_{T_1}(q^2) g^{\mu j} g^{\nu k}\varepsilon_{ijk}\sigma^i 
	              + (g^{\mu i} g^{\nu 0} - g^{\mu 0} g^{\nu i})T_i \\
	&
	\pm \frac{i}{2} \varepsilon^{\mu\nu\rho\sigma} \left[ (g_{\rho i}g_{\sigma 0} 
	- g_{\rho 0} g_{\sigma i}) T^i + F_{T_1}(q^2) g_{\rho m} g_{\sigma n} \varepsilon^{mni}\sigma_i \right] + \dots, \nn
\end{align}
where
\begin{align} \label{eq:tbilinearterm}
	T^i &= \frac{i}{2m_p}\left[\left(F_{T_1}(q^2) - 2 F_{T_2}(q^2) \right)q^i I
	+ F_{T_1}(q^2) (\boldsymbol{\sigma}_a \times \mathbf{Q})^i \right].
\end{align}
Terms containing the momentum sum $\vecl{Q}$ are called recoil terms \cite{Tomoda:1990rs}.

The nuclear currents can be obtained from Eqs. \eqref{eq:sbilinear}-\eqref{eq:tbilinearterm} by summing over all neutrons, located at positions $\vecl{r}_a$, in the initial nucleus as
\begin{align}
	\mathcal{J}_K^\Xi(\vecl{x}) = \sum_a\tau^a_+\delta(\vecl{x}-\vecl{r}_a) J_{K,a}^\Xi,
\end{align}
where $J_{K,a}^\Xi$ denotes any of the nucleon currents\footnote{For positron emission, $\tau_+$ is replaced by $\tau_-$ and the sum is over protons.}.

The short-range $\ovbb$ decay transition involves two such currents, each transforming one neutron into a proton. Having the five different terms in the effective Lagrangian Eq.~\eqref{eq:lagsr} we thus need to evaluate five different products of nucleon currents in the non-relativistic expansion. Furthermore, we need to sum over all neutrons in the initial nucleus and the corresponding nuclear transition operators can be expressed as 
\begin{align}
\label{eq:transitionop}
	\mathcal{H}_K(\vecl{x},\vecl{y}) = \sum_{a\neq b}\tau_+^a \tau_+^b \delta(\vecl{x}-\vecl{r}_a) \delta(\vecl{y}-\vecl{r}_b) \Pi_{K,ab}^\Xi,
\end{align}
with the products of the relevant nucleon currents $\Pi_{K,ab}^\Xi$ given in Appendix~\ref{sec:currentproducts}. In the case of the terms 1, 2 and 3, the nuclear transition operator has no free Lorentz index whereas for terms 4 and 5, there remains one free index that is contracted with that of the electron current.

\subsection{From Nucleons to the Nucleus}
\label{sec:nucleonstonucleus}
The final and most challenging step in the determination of the $\ovbb$ NMEs concerns the calculation of the matrix elements at the \emph{nuclear} level. This requires an understanding of nuclear structure and given the highly complex nature of the many-body problem, it is not solvable from first principle. Above, we have constructed the short-range nuclear $\ovbb$ transition operators using the general nucleon operator with their $\mathbf{q}^2$ and thus distance dependence parametrized by experimentally constrained form functions. We now define the nuclear matrix elements as
\begin{align}
	\nme_K \equiv 
	\bra{\mathcal{O}^+_F} \mathcal{H}_K 
	\ket{\mathcal{O}^+_I},
\end{align}
with the transition operator given in Eq.~\eqref{eq:transitionop} and where $\bra{\mathcal{O}^+_F}$ and $\bra{\mathcal{O}^+_I}$ denote the wavefunctions of the final and initial nuclear state under consideration, respectively. In principle, we would also need the wavefunctions of the intermediate states formed by a single beta decay-like transition from one nucleon current. Exploiting the completeness of all intermediate states, we instead use the \emph{closure approximation} in directly calculating the above matrix element. This approximation is very well justified in the case of short-range operators as the intermediate transition occurs at very high energies $|q| \approx 100$~MeV, corresponding to the inter-nucleon distance, compared to the nuclear transition itself at $Q_{\beta\beta} \approx 1$~MeV. 

We concentrate on $0^+ \to 0^+$ transitions in this paper. In this case, all terms containing an odd number of $\vecs{\sigma}$ and/or an odd number of $\vecl{q}$, $\vecl{Q}$ occurrences in $\Pi_i$ vanish, due to angular momentum and parity selection rules, when only $S_{1/2}-S_{1/2}$ wave approximation of the electron wave functions is assumed.  Using the results in Appendix~\ref{sec:currentproducts}, the matrix elements for the five short-range operators can be collected. In each case we keep track of signs corresponding to different combinations of chiralities. For the first three operators (i.e. those proportional to $\epsilon_1$, $\epsilon_2$ and $\epsilon_3$) three sign possibilities are presented and they correspond to the following combinations of chiralities (in this order): $RR$, $LL$ and $(1/2)\nba{RL + LR}$. For the fourth and fifth operator (those proportional to $\epsilon_4$ and $\epsilon_5$) a row of four signs is shown, as in those cases the two hadronic currents have different Lorentz structures, thus all four possible combinations of chiralities have to be considered (in this order): $RR$, $LL$, $RL$ and $LR$. To keep the expressions simple, when all three/four signs are the same, we show only a single sign. Using this notation the matrix elements for the five different short-range operators read
\begin{align}
	\label{eq:nme1}
	\nme_1 =~
	&g_S^2\nme_F \\
	&\signs{+}{+}{-} \frac{g_{PS}^2}{12m_p^2}\left(\nme_{GT}^{'PP}-\nme_T^{'PP}\right),
	\nn\\
	\nme_2 =~
	&-2g_{T_1}^2\nme_{GT},\\
	\nme_3 =~
	&g_V^2 \nme_{F} \\
	&\signs{-}{-}{+} g_A^2 \nme_{GT}^{AA} \nn \\
	&\signs{+}{+}{-} \frac{g_A g_P}{6 m_p^2} \nba{\nme^{\prime AP}_{GT} - \nme^{\prime AP}_{T}} \nn \\
	&+ \frac{\nba{g_V+g_{W}}^2}{12m_p^2}\nba{\nme^{\prime}_{GT} + \frac{1}{2} \nme^{\prime}_{T}} \nn \\
	&\signs{-}{-}+{} \frac{g_P^2}{24 m_p^4} \nba{\nme^{\prime\prime PP}_{GT} - \nme^{\prime\prime PP}_{T}}, \nn \\
	\nme_4^\mu =~& \foursigns{-}{-}{+}{+} i g^{\mu 0} g_A g_{T_1} \nme_{GT}^A \\
	&\foursigns{+}{+}{-}{-} ig^{\mu 0}\frac{g_Pg_{T_1}}{12m_p^2}\nba{\nme^{\prime P}_{GT} - \nme^{\prime P}_{T}}, \nn \\
\label{eq:nme5}
	\nme_5^\mu =~& g^{\mu 0} g_S g_V\nme_F   \\
	&\foursigns{+}{+}{-}{-} g^{\mu 0} \frac{g_Ag_{PS}}{12m_p^2} \nba{\tilde{\nme}^{AP}_{GT} - \tilde{\nme}^{AP}_{T}} \nn \\
	&\foursigns{-}{-}{+}{+} g^{\mu 0} \frac{g_Pg_{PS}}{24m_p^3}\nba{\nme^{\prime q_0 PP}_{GT} - \nme^{\prime q_0 PP}_{T}}. \nn 
\end{align}
In these expressions, we have kept all terms to order $1$. In $\nme_3$ we have retained also the term proportional to $\nba{F_V(q^2)+F_{W}(q^2)}^2$, which is a bit smaller, but may still represent an important contribution of the respective operator. We have also separated out from the form factors $F_X(q^2)$, the so-called charges, i.e. the values at $q^2 = 0$, $F_X(0)\equiv g_X$. The $q$-dependence is then given by (here $X \in \{S,V,W,T_1,T_2,T_3\}$)
\begin{align}
	\tilde{h}(q^2) &= \frac{1}{\left(1+q^2/m_V^2\right)^4}.
\end{align}
We treat separately the $A$, $P$ and $PS$ couplings which have a different $q$-dependence. If the axial-vector coupling $A$ is present in the first or second power, the $q$-dependence reads
\begin{align}
	\tilde{h}_A(q^2) &= \frac{1}{\left(1+q^2/m_V^2\right)^2} \frac{1}{\left(1+q^2/m_A^2\right)^2}, \\
	\tilde{h}_{AA}(q^2) &= \frac{1}{\left(1+q^2/m_A^2\right)^4},
\end{align}
respectively.

Similarly, if a single power of pseudoscalar coupling in combination with the axial-vector coupling $A$ or some other coupling $X$ is present, then we have
\begin{align}
	\tilde{h}_{AP}(q^2) &= 
	\frac{1}{\left(1 + q^2/m_A^2\right)^4}\frac{1}{\left(1 + q^2/m_\pi^2\right)}, \\
	\tilde{h}_{P}(q^2) &= 
	\frac{1}{\left(1 + q^2/m_V^2\right)^4}\frac{1}{\left(1 + q^2/m_\pi^2\right)},
\end{align}
respectively, while in case of the second power of pseudoscalar coupling the $q$-dependence has the form
\begin{align}
	\tilde{h}_{PP}(q^2) = 
	\frac{1}{\left(1 + q^2/m_V^2\right)^4}\frac{1}{\left(1 + q^2/m_\pi^2\right)^2}.
\end{align}

The Fermi ($F$), Gamow-Teller ($GT$) and tensor ($T$) matrix elements appearing in Eqs.~\eqref{eq:nme1}-\eqref{eq:nme5} can be calculated in any nuclear structure model \cite{Simkovic:2007vu, Caurier:2007wq, Barea:2013bz}. We follow in this article the formulation of \cite{Simkovic:1999re} and \cite{Barea:2013bz} where the two-body transition operator $H$ is constructed in momentum space as the product of the so-called neutrino potential, $v(q)$, times the form factors $\tilde{h}(q^2)$. Since we consider short-range mechanisms with a $\delta$-function in configuration space, the Fourier transform is a constant, and the neutrino potential in momentum space is \cite{Barea:2013bz, Simkovic:1999re}
\begin{align}
\label{eq:neutrinopotential}
	v(q)=\frac{2}{\pi}\frac{1}{m_e m_p},
\end{align}
where we have used the standard normalization. Incidentally, for the long-range mechanism the neutrino potential is
\begin{align}
	v(q) = \frac{2}{\pi}\frac{1}{q(q+\tilde{A})},
\end{align}
where $\tilde{A}$ is the closure energy. This formulation allows one therefore to calculate simultaneously all matrix elements, short- and long-range, by simply specifying the neutrino potential.

As a further aside, to do calculations in coordinate space, one simply takes the Fourier-Bessel transforms of the product of the neutrino potential $v$ times the form factor $\tilde{h}$,
\begin{align}
	h(r) = \frac{2}{\pi} \int_0^\infty j_\lambda(q^2)
	\frac{1}{m_e m_p} \tilde{h}(q) q^2 \mathrm{d}q,
\end{align}
where $\lambda = 0$ for Fermi and Gamow-Teller contributions and $\lambda=2$ for a tensor contribution.

Finally, an additional improvement is the introduction of short-range correlations (SRC) in the nuclear structure calculation. These are of crucial importance for short-range non-standard mechanisms and they can be taken into account by multiplying the potential $v(r)$ in coordinate space by a correlation function $f(r)$ squared. The most commonly used correlation function is the Jastrow function,
\begin{align}
	f_J(r) = 1 - ce^{-ar^2}(1 - br^2)
\end{align}
with $a = 1.1~\text{fm}^{-2}$, $b = 0.68~\text{fm}^{-2}$ and $c = 1$ for the phenomenological Miller-Spencer parametrization \cite{MILLER1976562}, and $a = 1.59~\text{fm}^{-2}$, $b = 1.45~\text{fm}^{-2}$ and $c = 0.92$ for the Argonne parametrization \cite{Simkovic:2009pp}. Since the formulation described above is in momentum space, we take SRC's into account by using the Fourier-Bessel transform of $f_J(r)$. 

Introducing
\begin{align}
	h_{\circ}(q^2) = \frac{2}{\pi} \frac{1}{m_e m_p} \tilde{h}_{\circ}(q^2),
\end{align}
(where the placeholder $\circ$ is used to note that the same redefinition is used for all the above defined types of $q$-dependencies) and the notation
\begin{align}
	\langle H_{ab} \rangle = \bra{\mathcal{O}^+_F} \sum_{a\neq b}\tau^+_a \tau^+_b H_{ab} \ket{\mathcal{O}^+_I},
\end{align}
where $H_{ab}$ denotes any two-body operator, we can write the Fermi, $\mathcal{M}_F$, and Gamow-Teller, $\mathcal{M}_{GT}$, matrix elements appearing in Eqs. \eqref{eq:nme1}-\eqref{eq:nme5} as
\begin{align}
	\mathcal{M}_F    &= \langle h(q^2) \rangle, \\
	\mathcal{M}_{GT} &= \langle h(q^2) (\vecs{\sigma}_a\cdot\vecs{\sigma}_b) \rangle.
\end{align} 

When the Gamow-Teller matrix element comes with one or two powers of the axial-vector coupling, we define
\begin{align}
	\mathcal{M}_{GT}^A &= \langle h_A(q^2) (\vecs{\sigma}_a\cdot\vecs{\sigma}_b) \rangle, \\
	\mathcal{M}_{GT}^{AA} &= \langle h_{AA}(q^2) (\vecs{\sigma}_a\cdot\vecs{\sigma}_b) \rangle.
\end{align} 

In the third short-range operator, matrix elements $\mathcal{M}^{\prime}_{GT}$ and $\mathcal{M}^{\prime}_T$ appear, which are defined as
\begin{align}
	\frac{1}{m_p^2}\mathcal{M}^{\prime}_{GT} &= \left\langle \frac{\vecl{q}^2}{m_p^2} h(q^2) (\vecs{\sigma}_a\cdot\vecs{\sigma}_b) \right\rangle, \\
	\frac{1}{m_p^2}\mathcal{M}^{\prime}_{T} &= \left\langle \frac{1}{m_p^2} \sqb{\vecl{q}^2-\frac{1}{3}(\vecl{q}\cdot \hat{\vecl{r}}_{ab})^2} h(q^2) \vecl{S}_{ab} \right\rangle.
\end{align}
Since $q \sim 100$~MeV in $\ovbb$ decay, these terms are suppressed by a factor of $\mathcal{O}(0.01)$ relative to the standard terms $\mathcal{M}_{GT}$ and $\mathcal{M}_{T}$. However, the enhancement of the corresponding form factor partly compensates this; therefore, we include them. These matrix elements can be easily calculated since the neutrino potential is just a function of $\vecl{q}^2$.

Similarly, the matrix elements $\mathcal{M}^{\prime P}_{GT}$, $\mathcal{M}^{\prime P}_T$ and $\mathcal{M}^{\prime AP}_{GT}$, $\mathcal{M}^{\prime AP}_T$ are given by
\begin{align}
	\frac{1}{m_p^2}\mathcal{M}^{\prime P}_{GT} &= \left\langle \frac{\vecl{q}^2}{m_p^2} h_{P}(q^2) (\vecs{\sigma}_a\cdot\vecs{\sigma}_b) \right\rangle, \\
	\frac{1}{m_p^2}\mathcal{M}^{\prime P}_{T} &= \left\langle \frac{1}{m_p^2} \sqb{\vecl{q}^2-\frac{1}{3}(\vecl{q}\cdot \hat{\vecl{r}}_{ab})^2} h_{P}(q^2) \vecl{S}_{ab} \right\rangle,
\end{align} 
and
\begin{align}
	\frac{1}{m_p^2}\mathcal{M}^{\prime AP}_{GT} &= \left\langle \frac{\vecl{q}^2}{m_p^2} h_{AP}(q^2) (\vecs{\sigma}_a\cdot\vecs{\sigma}_b) \right\rangle, \\
	\frac{1}{m_p^2}\mathcal{M}^{\prime AP}_{T} &= \left\langle \frac{1}{m_p^2} \sqb{\vecl{q}^2-\frac{1}{3}(\vecl{q}\cdot \hat{\vecl{r}}_{ab})^2} h_{AP}(q^2) \vecl{S}_{ab} \right\rangle,
\end{align} 
respectively. These terms are also smaller by a factor of $\mathcal{O}(0.01)$ relative to the standard terms $\mathcal{M}_{GT}$ and $\mathcal{M}_{T}$. Nonetheless, this suppression is compensated by the enhancement of the $g_P$ form factor.

Next, we have terms $\mathcal{M}_{GT}^{'PP}$, $\mathcal{M}_{T}^{'PP}$ contributing to the first short-range operator, which read
\begin{align} \label{eq:PSNMEGT}
\frac{1}{m_p^2}\mathcal{M}_{GT}^{'PP} &= \left\langle \frac{\vecl{q}^2}{m_p^2} h_{PP}(q^2) (\vecs{\sigma}_a\cdot\vecs{\sigma}_b) \right\rangle, \\
\label{eq:PSNMET}
\frac{1}{m_p^2}\mathcal{M}_{T}^{'PP} &= \left\langle \frac{1}{m_p^2} \sqb{\vecl{q}^2-\frac{1}{3}(\vecl{q}\cdot \hat{\vecl{r}}_{ab})^2} h_{PP}(q^2) \vecl{S}_{ab} \right\rangle.
\end{align} 
These matrix elements are smaller by a factor of $\mathcal{O}(0.01)$. However, if the pseudoscalar coupling $g_{PS}$ is larger by two orders of magnitude as claimed in \cite{Gonzalez-Alonso:2018omy}, these terms become comparable to those with the Fermi and Gamow-Teller matrix elements $\mathcal{M}_F$ and $\mathcal{M}_{GT}$, or even larger.

The matrix elements $\mathcal{M}^{\prime\prime PP}_{GT}$ and $\mathcal{M}^{\prime\prime PP}_T$ appearing in the third short-range operator can be written as
\begin{align}
	\frac{1}{m_p^4}\mathcal{M}^{\prime\prime PP}_{GT} &= \left\langle \frac{\vecl{q}^4}{m_p^4} h_{PP}(q^2) (\vecs{\sigma}_a\cdot\vecs{\sigma}_b) \right\rangle, \\
	\frac{1}{m_p^4}\mathcal{M}^{\prime\prime PP}_{T} &= \left\langle \frac{\vecl{q}^2}{m_p^4} \sqb{\vecl{q}^2-\frac{1}{3}(\vecl{q}\cdot \hat{\vecl{r}}_{ab})^2} h_{PP}(q^2) \vecl{S}_{ab} \right\rangle.
\end{align} 
Again, since $q \sim 100$~MeV in $\ovbb$ decay, these terms are smaller by a factor of $\mathcal{O}(10^{-4})$ relative to the standard terms $\mathcal{M}_{GT}$ and $\mathcal{M}_{T}$. However, this suppression is again balanced by the enhancement of the form factor $g_P$, which appears here in the second power. These terms can be easily calculated since the neutrino potential is just a function of $\vecl{q}^2$.

The terms $\tilde{\nme}^{AP}_{GT}$ and $\tilde{\nme}^{AP}_T$, also called recoil terms, are defined as
\begin{align}
\frac{1}{m_p^2}\tilde{\mathcal{M}}^{AP}_{GT} &= \left\langle \frac{\vecl{Q}\cdot\vecl{q}}{m_p^2} h_{AP}(q^2) (\vecs{\sigma}_a\cdot\vecs{\sigma}_b) \right\rangle, \\
\frac{1}{m_p^2}\tilde{\mathcal{M}}^{AP}_{T} &= \left\langle \frac{1}{m_p^2} \sqb{\vecl{Q}\cdot\vecl{q}-\frac{1}{3}(\vecl{Q}\cdot \hat{\vecl{r}}_{ab})(\vecl{q}\cdot \hat{\vecl{r}}_{ab})} h_{AP}(q^2) \vecl{S}_{ab} \right\rangle.
\end{align} 
Although these terms are suppressed by a factor of $\mathcal{O}(0.01)$, if the pseudoscalar coupling $g_{PS}$ is larger by two orders of magnitude, they will become important. These matrix elements are difficult to calculate since the operator $\vecl{Q}$ is not simply a function of $\vecl{q}^2$. A good estimate can be however obtained by replacing $(\vecl{Q}\cdot\vecl{q}) / m_p^2$ with the expectation value in the state $\ket{O^+_I}$, $\left\langle \vecl{Q}\cdot\vecl{q} / m_p^2 \right\rangle \sim 0.01$, in which case
\begin{align}
	\frac{1}{m_p^2}\tilde{\nme}_{GT}^{AP} &= \left\langle \frac{\vecl{Q}\cdot\vecl{q}}{m_p^2} \right\rangle \mathcal{M}_{GT}^{AP},
\end{align}
etc.

Finally, the terms $\mathcal{M}_{GT}^{\prime q_0 PP}$, $\mathcal{M}_{T}^{\prime q_0 PP}$ appearing in the fifth matrix element read
\begin{align}
\frac{1}{m_p^3}\mathcal{M}_{GT}^{\prime q_0 PP} &= \left\langle \frac{q_0 \vecl{q}^2}{m_p^3} h_{PP}(q^2) (\vecs{\sigma}_a\cdot\vecs{\sigma}_b) \right\rangle, \\
\frac{1}{m_p^3}\mathcal{M}_{T}^{\prime q_0 PP} &= \left\langle \frac{q_0}{m_p^3} \sqb{\vecl{q}^2-\frac{1}{3}(\vecl{q}\cdot \hat{\vecl{r}}_{ab})^2} h_{PP}(q^2) \vecl{S}_{ab} \right\rangle.
\end{align} 
Since $q \sim 100$~MeV and $q_0 \sim 10$~MeV in $\ovbb$ decay, these terms are smaller by a factor of $\mathcal{O}(10^{-4})$ relative to the terms $\mathcal{M}_{GT}$ and $\mathcal{M}_{T}$. However, this suppression is balanced by the enhancement of the form factors $g_P$ and $g_{PS}$.

The above basic building blocks are written in terms of the Pauli spin operators $\vecs{\sigma}$, the nucleon momenta difference $\vecl{q}$ and sum $\vecl{Q}$, and the direction unit vector between two nucleons, $\vecl{\hat r}_{ab} = \vecl{r}_{ab}/|\vecl{r}_{ab}|$. Furthermore, we use $\vecl{S}_{ab} = 3(\vecs{\sigma}_a \cdot \vecl{\op{r}}_{ab}) (\vecs{\sigma}_b \cdot \vecl{\op{r}}_{ab}) - (\vecs{\sigma}_a \cdot \vecs{\sigma}_b)$.

Let us remark that in case of the short-range operator $\mathcal{O}_1$ incorporating scalar and pseudoscalar quark currents, the enhancement of the pseudoscalar form factor $g_{PS}$ can make the third-order term of the non-relativistic expansion of the pseudoscalar current important. We anticipate this term to be of the order $F_{PS}(q^2)\mathcal{O}(\vecl{q}^3/m_p^3)$ and its product with the first-order pseudoscalar term of the expansion would give a contribution $F_{PS}^2(q^2) \mathcal{O}(\vecl{q}^4/m_p^4) \sim \mathcal{O}(1)$. This contribution is not listed in the above paragraphs, because only terms up to the order of $\vecl{q}/m_p$ in the currents are considered in this paper. However, we conjecture that it will always be sub-dominant to the terms in Eqs.~\eqref{eq:PSNMEGT} and \eqref{eq:PSNMET}, which are larger by two orders of magnitude. The exact relative size of these contributions depends, of course, on the actual size of the corresponding NMEs, but there is no reason to believe that the NME involving the third-order term of the expansion should be exceptionally large.

\section{Leptonic Phase Space Factors}
\label{sec:PSFs}
The leptonic phase-space factors describe the atomic part of the physics involved in $\ovbb$ decay. They quantify the effect of the relativistic electrons emitted in the process. The position-dependent wavefunction of each electron can be expanded in terms of spherical waves,
\begin{align}
	e_{\vecl{p}s}(\vecl{r}) = 
	  e_{\vecl{p}s}^{S_{1/2}}(\vecl{r}) 
    + e_{\vecl{p}s}^{P_{1/2}}(\vecl{r}) + \dots,
\end{align}
where $\vecl{p}$ is the asymptotic momentum of the electron at long distance and $s$ denotes its spin projection. The $S_{1/2}$ and the $P_{1/2}$ waves on the right-hand side of the above expansion are respectively given by \cite{Tomoda:1990rs}
\begin{align} \label{eq:radwaves}
	e^{S_{1/2}}_{\vecl{p}s}(\vecl{r}) = 
	\begin{pmatrix}
		g_{-1}(E, r) \chi_s\\ 
		   f_1(E, r)(\vecs{\sigma}\cdot\vecl{\op p}) \chi_s 
	\end{pmatrix}, \quad
	e^{P_{1/2}}_{\vecl{p}s}(\vecl{r}) = 
	i\begin{pmatrix}
		g_1(E, r)(\vecs{\sigma}\cdot\vecl{\op r})
		(\vecs{\sigma}\cdot\vecl{\op p})\chi_s \\
	   -f_{-1}(E, r)(\vecs{\sigma}\cdot\vecl{\op r})\chi_s 
	\end{pmatrix},
\end{align}
where $g_\kappa(E, r)$ and $f_\kappa(E, r)$ are the radial wavefunctions of the `large' and `small' components. The electron energy at asymptotically large distances is $E = \sqrt{\vecl{p}^2 + m_e^2}$ and its spin state is described by the two-dimensional spinor $\chi_s$. The Pauli matrices $\vecs{\sigma}$ here operate in the electron spin space. The wavefunctions satisfy the asymptotic boundary condition
\begin{align}
	\begin{pmatrix}
		g_\kappa(E, r) \\ 
		f_\kappa(E, r)
	\end{pmatrix}
	\xrightarrow{r\to\infty}
	\frac{e^{-i\Delta_\kappa^c}}{pr}
	\begin{pmatrix}
		\sqrt{\frac{E+m_e}{2E}}\sin\nba{pr+y\ln(2pr)
			- \tfrac{1}{2}\pi l_\kappa + \Delta_\kappa^c} \\ 
		\sqrt{\frac{E-m_e}{2E}}\cos\nba{pr+y\ln(2pr)-\tfrac{1}{2}\pi l_{\kappa}+\Delta_{\kappa}^c}
	\end{pmatrix},
\end{align}
where $\kappa=\pm(j + \frac{1}{2})$, $l_{\kappa} = j\pm\frac{1}{2}$, $y = \alpha Z_F E/p$ and $\Delta^c_{\kappa}$ is a phase shift. Here, $p = |\vecl{p}|$, $\alpha$ is the fine structure constant, $j$ is the total angular momentum of the electron. Inside the nucleus, the radial wavefunctions Eq.~\eqref{eq:radwaves} can be expanded in $r$ approximated by the leading terms as
\begin{align}
	\begin{pmatrix}
		g_{-1}(E, r)\\ 
		f_1(E, r)
	\end{pmatrix}
	\approx
	\begin{pmatrix}
		A_{-1} \\ 
		A_{+1}
	\end{pmatrix}\!, \,\,
	\begin{pmatrix}
		g_1(E, r) \\
	   -f_{-1}(E, r)
	\end{pmatrix}
	\approx
	\begin{pmatrix}
		 A_{+1}\sqb{\tfrac{1}{2}\alpha Z_F+\tfrac{1}{3}\nba{E+m_e}R_A}\frac{r}{R_A} \\ 
		-A_{-1}\sqb{\tfrac{1}{2}\alpha Z_F+\tfrac{1}{3}\nba{E-m_e}R_A}\frac{r}{R_A}
	\end{pmatrix}\!,
\end{align}
for $S_{1/2}$ and $P_{1/2}$ waves, respectively. Here $A_\kappa$ are normalization constants and $R_A$ is radius of the daughter nucleus. In the limit $Z_F\to 0$ the radial wavefunctions acquire the form of spherical Bessel functions, while the normalization constants become $A_{\pm 1}\approx \sqrt{(E\mp m_e)/(2E)}$. In our calculations, however, the above shown approximations are not employed, as we derive the phase-space factors using numerically calculated radial wavefunctions as described in \cite{Kotila:2012zza}. Therein a numerical solution is performed using a piecewise exact power series expansion of the radial wave functions. On top of the Coulomb potential of the daughter nucleus (with charge $Z_F$), $V(r) = -\alpha Z_F/r$, nuclear size and electron cloud screening corrections are taken into account. As a result, the considered potential reads
\begin{align}
	V(r) = 
	\begin{cases}
		- \alpha Z_F\frac{3-(r/R_A)^2}{2R_A} \times \varphi(r), &r < R, \\
		- \frac{\alpha Z_F}{r} \times \varphi(r),               &r \geq R,
	\end{cases}
\end{align}
where $\varphi(r)$ is the Thomas-Fermi function taking into account the electron screening. The non-trivial $r$-dependence of the above potential for $r < R$ is a result of the finite nuclear size, when a uniform charge distribution in a sphere of radius $R_A = R_0 A^{1/2}$ with $R_0 = 1.2$~fm is considered. In order to calculate the electron currents involved in $\ovbb$ decay we in principle need to evaluate the wavefunctions at the position of the corresponding transition. To be exact, this would require the wavefunction of the nucleon undergoing the respective decay, ideally from the nuclear structure model, or using a simplified harmonic oscillator wavefunctions \cite{Kotila:2012zza}. 

Instead, we follow \cite{Kotila:2012zza} and adopt the approximation of evaluating the electron wavefunction at the nuclear radius $r = R_A$,
\begin{align}
	f_{\pm 1}(E) \equiv f_{\pm 1}(E,R_A),\quad g_{\pm 1}(E) \equiv g_{\pm 1}(E,R_A).
\end{align}
This choice reflects the fact that nucleons largely decay at the surface of the nucleus due to Pauli-blocking of inner states.

For $0^+ \to 0^+$ transitions, parity-even nucleon operators need to be combined with $S_{1/2} - S_{1/2}$ and $P_{1/2} - P_{1/2}$ electron wave functions, while parity-odd operators need to be combined with $S_{1/2} - P_{1/2}$ wave functions. The calculation of the leptonic squared matrix elements are outlined in Appendix~\ref{sec:leptonicmatrixelements} and the results for $S_{1/2} - S_{1/2}$ wavefunctions are
\begin{align} 
\label{eq:PSFs1}
	\sum_{s_1,s_2}(\bar{e}_1(1+\gamma_5)e_2^c)
	(\bar{e}_1(1\pm\gamma_5)e_2^c)^\dagger
	\frac{1-P_{e_1 e_2}}{2} &= 
	2\left[f_{11}^{(0)} + f_{11\pm}^{(1)}(\hat{\vecl{p}}_1\cdot\hat{\vecl{p}}_2) \right]\!, \\
	\sum_{s_1,s_2}(\bar{e}_1\gamma_\mu \gamma_5 e_2^c)
	(\bar{e}_1\gamma_\nu \gamma_5 e_2^c)^\dagger 	\frac{1-P_{e_1e_2}}{2} &= 
	\frac{1}{8}\left[f^{(0)}_{66} 
	+ f^{(1)}_{66}(\vecl{\op p}_1\cdot\vecl{\op p}_2) \right]\!, 
	\, (\mu,\nu=0), 
	\label{eq:PSFs3}\\
	\sum_{s_1,s_2}(\bar{e}_1\gamma_\mu\gamma_5 e_2^c)
	(\bar{e}_1(1\pm\gamma_5)e_2^c)^\dagger \frac{1-P_{e_1e_2}}{2} &= 
	\mp \frac{1}{2}f_{16}^{(0)}, \quad\qquad\qquad\qquad (\mu=0). 
\label{eq:PSFs4} 
\end{align}
Here, $\vecl{\op p}_1\cdot\vecl{\op p}_2 = \cos\theta$ is the scalar product between the asymptotic momentum vectors of the two electrons, yielding the opening angle $0 \leq \theta \leq \pi$. In Eq.~\eqref{eq:PSFs1}, if both the involved scalar currents are left-handed, the same result as for two right-handed currents holds. The quantities $f^{(0)}_{ij} = f^{(0)}_{ij}(E_1,E_2)$ and $f^{(1)}_{ij} = f^{(1)}_{ij}(E_1,E_2)$ are given by
\begin{align}
\label{eq:PSFscal}
	f_{11}^{(0)}    &=  |f^{-1-1}|^2+|f_{11}|^2+|f{^{-1}}_1|^2+|{f_1}^{-1}|^2, &
	f_{11\pm}^{(1)} &= -2\left[{f^{-1}}_1{f_1}^{-1}\pm f^{-1-1}f_{11}\right], \\
	f_{66}^{(0)} &= 16\left[|f^{-1-1}|^2+|f_{11}|^2\right], &
	f_{66}^{(1)} &= 32\left[f^{-1-1}f_{11}\right], \label{eq:PSFcal2}\\
	f_{16}^{(0)} &=  4\left[|f_{11}|^2-|f^{-1-1}|^2\right], &
	f_{16}^{(1)} &=  0.
\label{eq:PSFcal3}
\end{align}
We note that our results agree with those of P\"{a}s et al. \cite{Pas:2000vn} and Tomoda \cite{Tomoda:1990rs}, except for the extra interference term $f_{11-}^{(1)}$ in Eq.~\eqref{eq:PSFscal} between the left- and right-handed scalar electron currents, and the fact that these authors use the notation of Doi \cite{Doi:1981, Doi:1983}, while we have used that of Tomoda \cite{Tomoda:1990rs}. The phase space factors corresponding to $\mu = j$ or $\nu = j$ in Eqs.~\eqref{eq:PSFs3} and \eqref{eq:PSFs4} are not shown, as their corresponding contributions to $\ovbb$ decay do not trigger $0^+ \rightarrow 0^+$ transition, in the case of $S_{1/2} - S_{1/2}$ approximation, we are interested in (although they are relevant when general $0^+ \to J^+$ transitions are considered). All the above listed phase space factors are given in terms of the underlying energy-dependent wave functions of the two electrons
\begin{align}
	f^{-1-1}   &= g_{-1}(E_1)g_{-1}(E_2), \\
	f_{11}     &= f_1   (E_1)f_1(E_2),    \\
	{f^{-1}}_1 &= g_{-1}(E_1)f_1(E_2),    \\
	{f_1}^{-1} &= f_1   (E_1)g_{-1}(E_2).
\end{align}
%

\section{Numerical Results}
\label{sec:numresults}

\subsection{Decay Half-life and Angular Correlation}

Combining the above results, the coefficients $a(E_1)$ and $b(E_1)$ in the fully differential rate Eq.~\eqref{eq:differentialrate} for $0^+ \to 0^+$~$\ovbb$ decay are given by
\begin{align}
\label{eq:aterm}
	a(E_1) &= 2 f_{11}^{(0)}\left|\sum_{I=1}^3\epsilon_I\nme_I\right|^2 
			 \!+\! \frac{1}{8} f_{66}^{(0)}\left|\sum_{I=4}^5\epsilon_I\nme_I\right|^2 
			 \!\mp\!
			 f_{16}^{(0)}\text{Re}\left[\left(\sum_{I=1}^3\epsilon_I\nme_I\right) 
			 \!\!\left(\sum_{I=4}^5\epsilon_I\nme_I\right)^*\right], \\
	b(E_1) &= 2 f_{11\pm}^{(1)} \left|\sum_{I=1}^3 \epsilon_I\nme_I\right|_{\pm}^2 
			 + \frac{1}{8} f_{66}^{(1)}\left|\sum_{I=4}^5\epsilon_I\nme_I\right|^2.
\label{eq:bterm}
\end{align}
They are expressed in terms of the NMEs in Eqs.~\eqref{eq:nme1}-\eqref{eq:nme5} and the PSFs in Eqs.~\eqref{eq:PSFscal}-\eqref{eq:PSFcal3}, where the summations as indicated are over the different current types $i=1,2,3,4,5$ including their different chiralities, $I=(i,XYZ)$ with $X,Y,Z \in \{L,R\}$. In Eq.~\eqref{eq:aterm}, the sign in front of $f_{16}^{(0)}$ is negative (positive) if the chirality of the electron scalar current involved in the interference term is $R$ ($L$). In Eq.~\eqref{eq:bterm}, the $\pm$ in the subscript of the norm symbolically denotes that the terms containing $\epsilon_I$ and $\epsilon_J^*$ corresponding to the same electron chiralities are accompanied by $f_{11+}^{(1)}$, while terms with opposite electron chiralities, the respective PSF is given by $f_{11-}^{(1)}$. The contribution of the interference term between currents $i=1,2,3$ and $i=4,5$ vanishes in $b(E_1)$ due to $f_{16}^{(1)} = 0$ in Eq.~\eqref{eq:PSFcal3}.

\setlength{\tabcolsep}{0.40em}
\renewcommand{\arraystretch}{1.15}
\begin{table}[t!]
	\centering
	\begin{tabular}{c|rrr} \hline
		 & $\nme_F$ & $\nme_{GT}$ & $\nme_T$ 
		 \\\hline
		$\prescript{76}{}{\text{Ge}}$  & $-42.8$ & $104.0$ & $-26.9$ 
		 \\
		$\prescript{82}{}{\text{Se}}$  & $-37.1$ &  $87.2$ & $-27.3$ 
		 \\
		$\prescript{100}{}{\text{Mo}}$ & $-46.8$ & $111.0$ &  $24.2$ 
		 \\
		$\prescript{130}{}{\text{Te}}$ & $-37.9$ &  $84.8$ & $-16.6$ 
		 \\ 
		$\prescript{136}{}{\text{Xe}}$ & $-29.7$ &  $66.8$ & $-12.7$ 
		 \\
		\hline
	\end{tabular}
	\caption{Nuclear matrix elements $\nme_F$, $\nme_{GT}$ and $\nme_{T}$ for selected nuclei, adopted from \cite{Barea:2015kwa}.}
	\label{tab:nmes}
\end{table}
The basic nuclear matrix elements $\nme_F$, $\nme_{GT}$ and $\nme_T$ on which the NMEs $\nme_I$ are based are given in Tab.~\ref{tab:nmes} for selected nuclei. The values are taken from Tab.~IV of \cite{Barea:2015kwa}. These matrix elements are given in dimensionless units, that is they are multiplied by the mass number dependent radius $R_A = R_0 A^{1/3}$ of the nucleus where $R_0 = 1.2$~fm.

\begin{figure}[t!]
	\begin{minipage}[b]{.48\textwidth}
		\centering
		\includegraphics[width=1\textwidth]{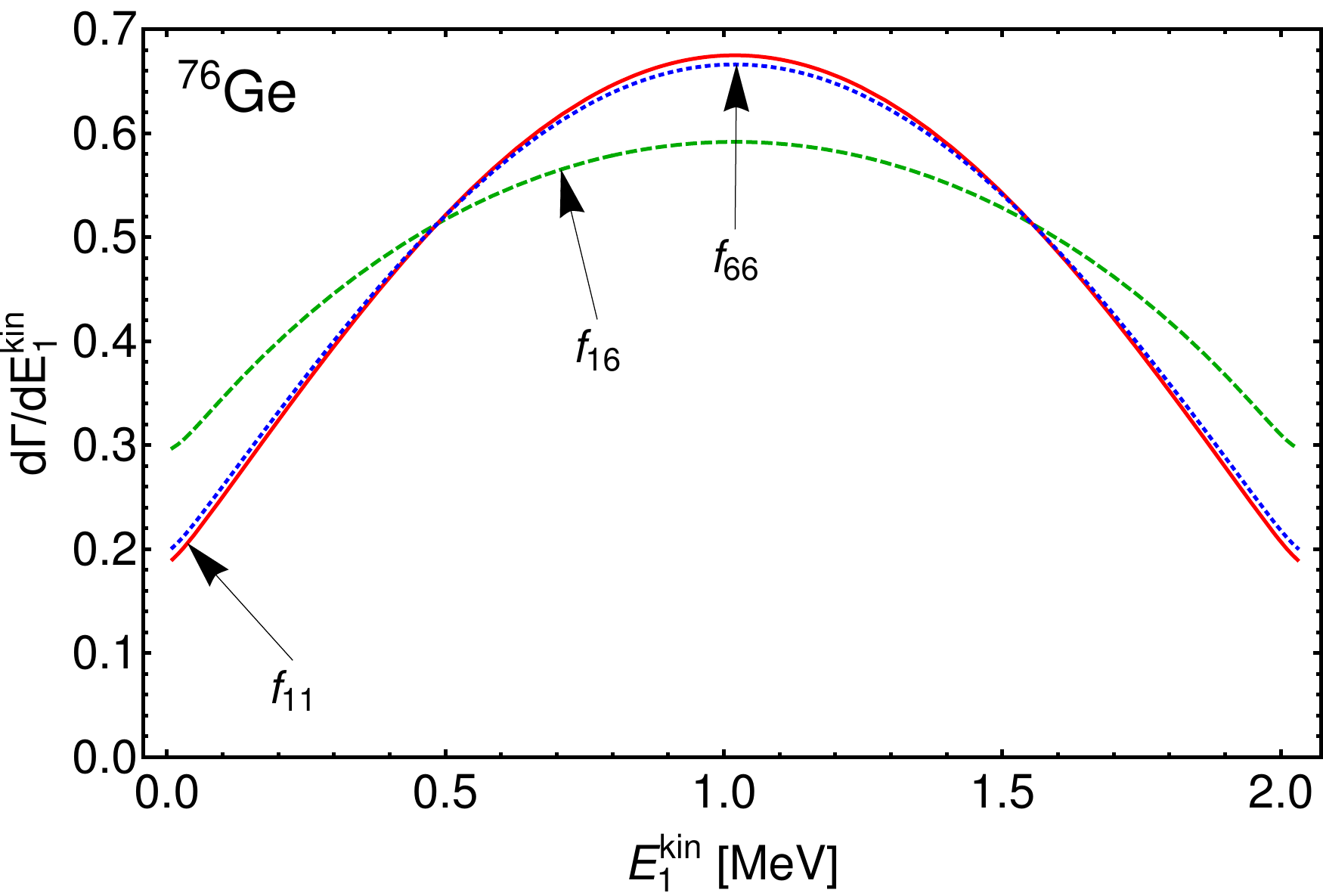}
	\end{minipage}
	\hfill
	\begin{minipage}[b]{.48\textwidth}
		\centering
		\includegraphics[width=1\textwidth]{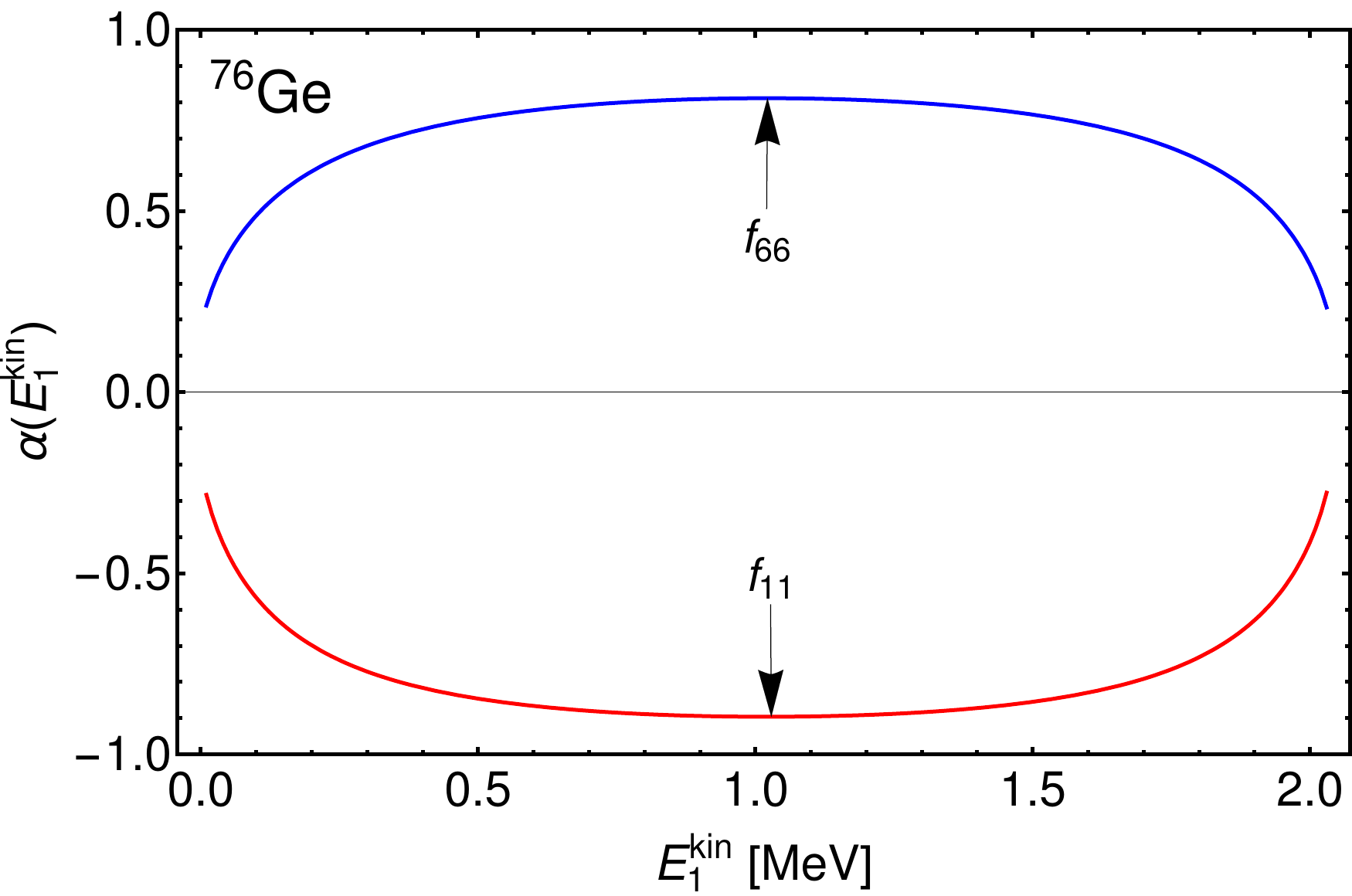}
	\end{minipage}
	\begin{minipage}[b]{.48\textwidth}
		\centering
		\includegraphics[width=1\textwidth]{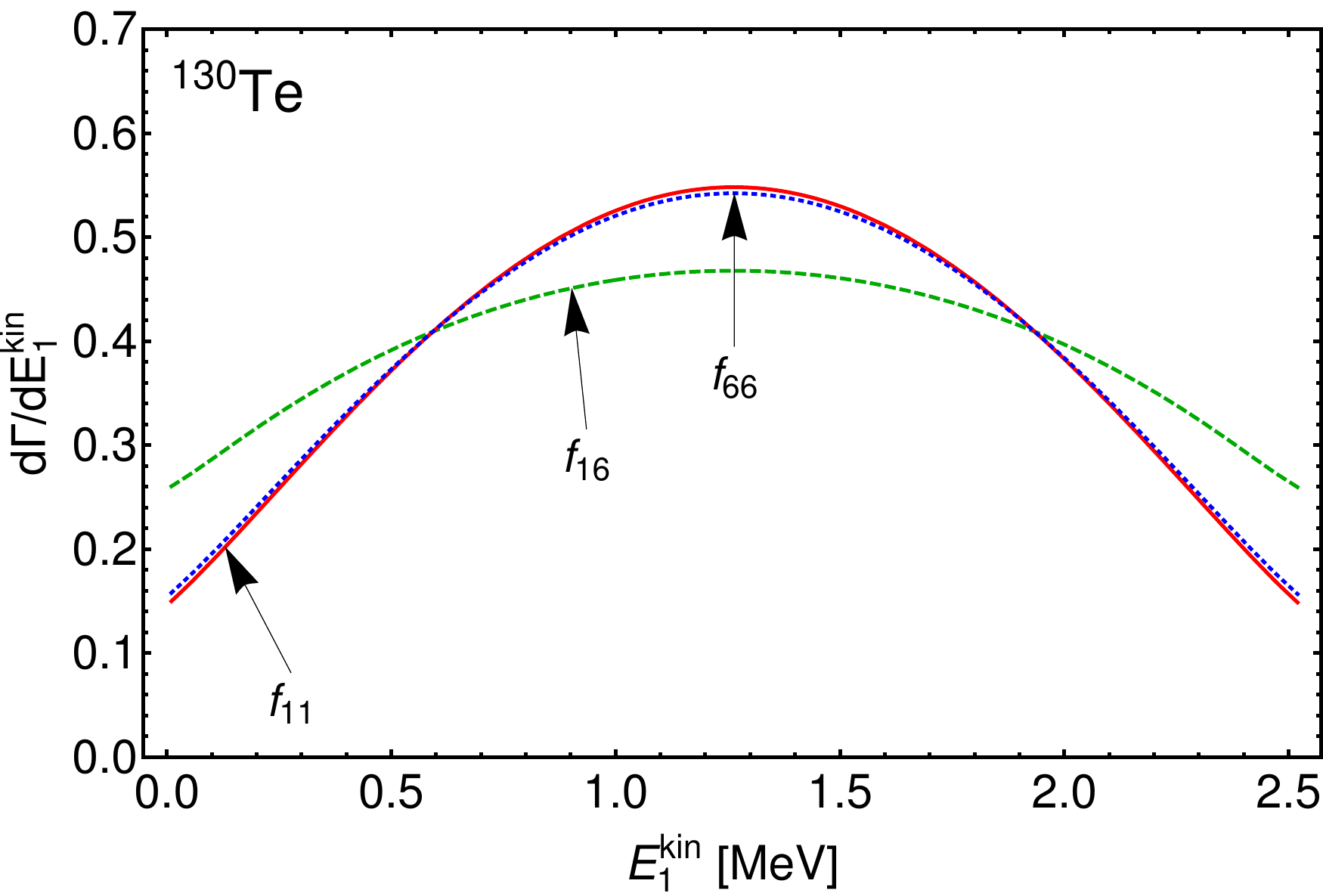}
	\end{minipage}
	\hfill
	\begin{minipage}[b]{.48\textwidth}
		\centering
		\includegraphics[width=1\textwidth]{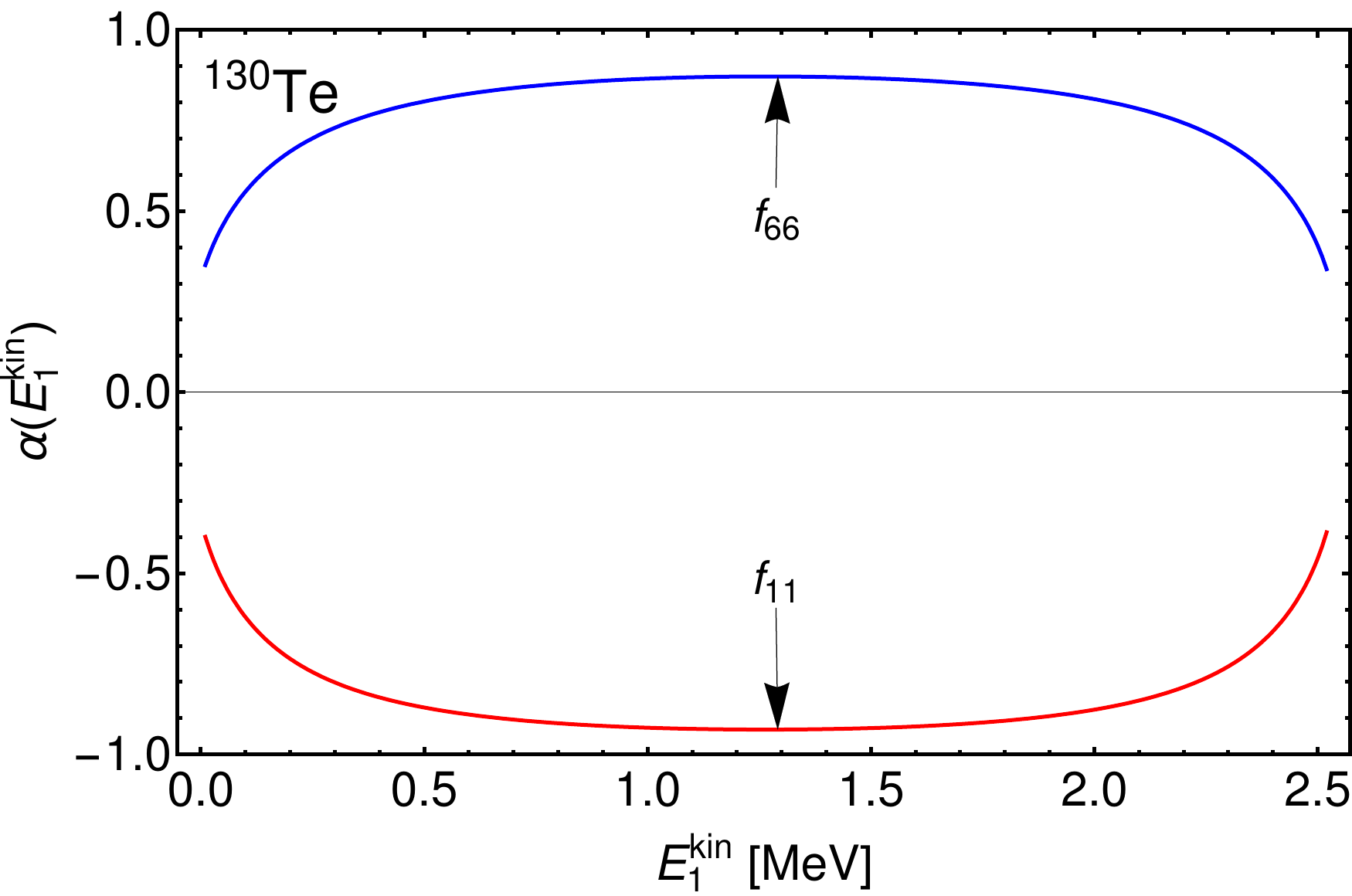}
	\end{minipage}
	\begin{minipage}[b]{.48\textwidth}
		\centering
		\includegraphics[width=1\textwidth]{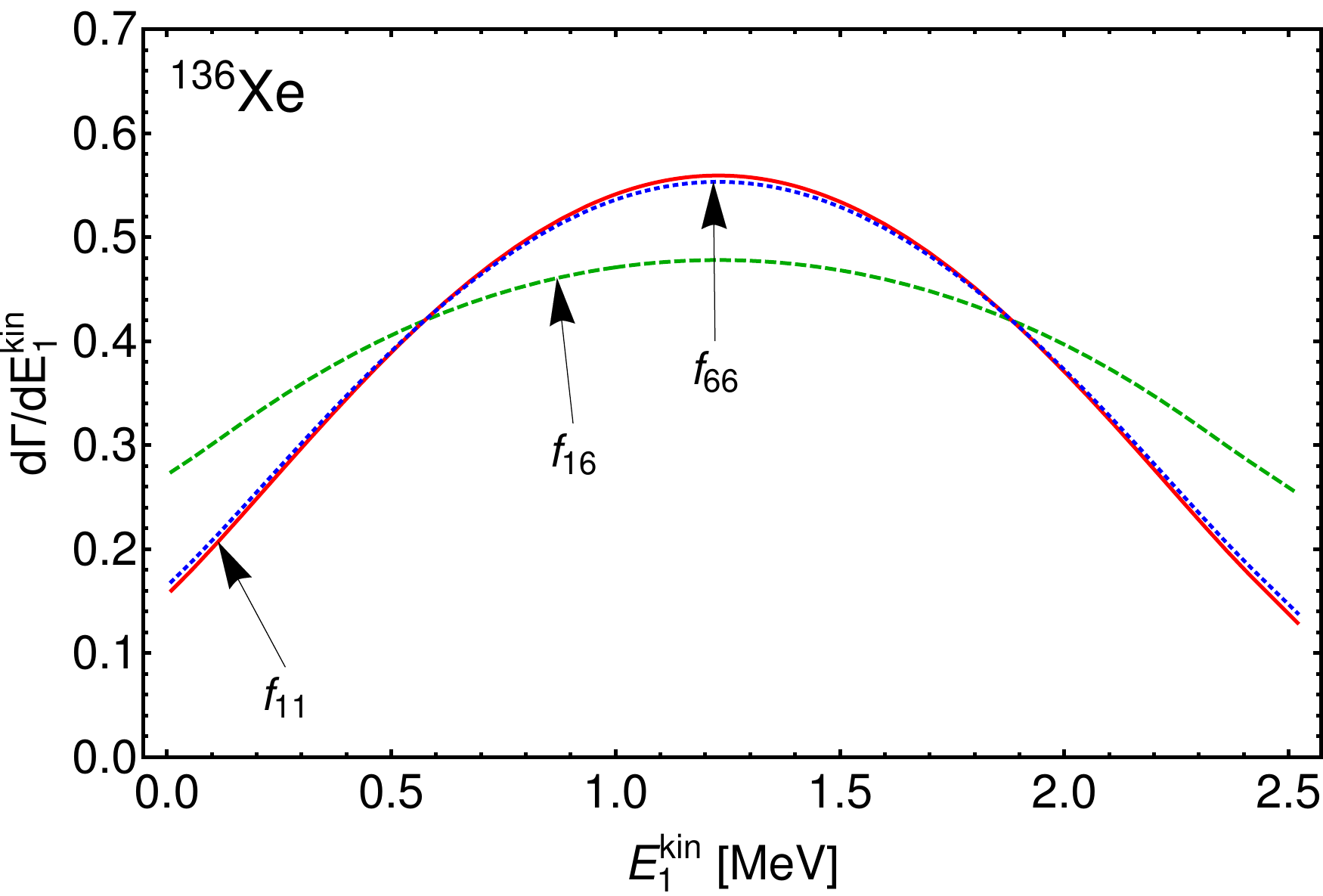}
	\end{minipage}
	\hfill
	\begin{minipage}[b]{.48\textwidth}
		\centering
		\includegraphics[width=1\textwidth]{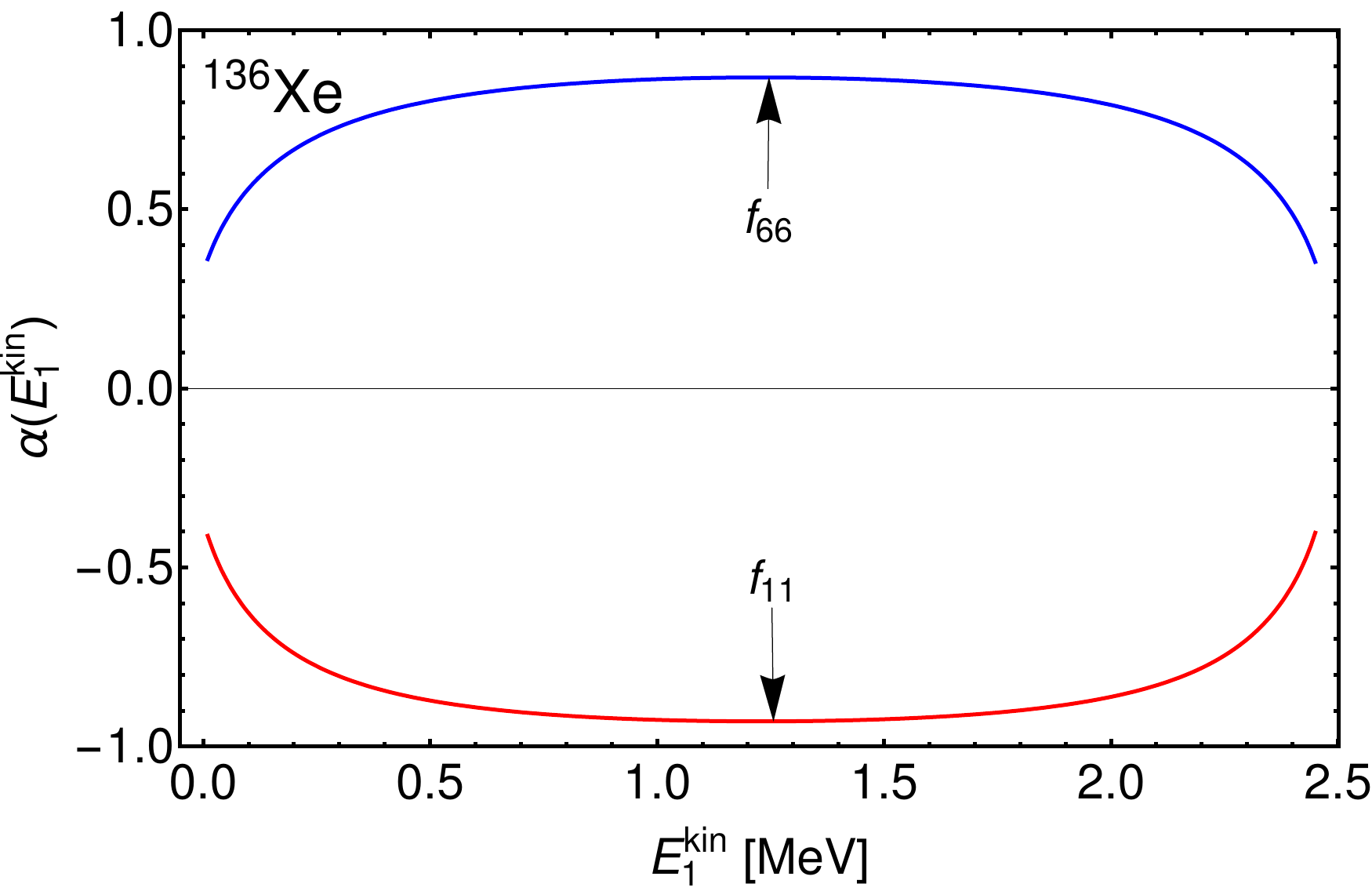}
	\end{minipage}
	\caption{Left panel: Single electron energy distribution $d\Gamma/dE^\text{kin}_1$ as function of the kinetic energy $E^\text{kin}_1=E_1-m_e$ for the three different phase space factors in Eq.~\eqref{eq:aterm}, namely $f_{11}$, $f_{66}$ and $f_{16}$. Right panel: Energy-dependent angular correlation $\alpha(E^\text{kin}_1)$ between the two electrons as function of the kinetic energy $E^\text{kin}_1$ for the phase space factors $f_{11}$ and $f_{66}$ (identically zero for $f_{16}$). From top to bottom, the plots show the results for the three $\ovbb$ decay isotopes $\prescript{76}{}{\text{Ge}}$, $\prescript{130}{}{\text{Te}}$ and $\prescript{136}{}{\text{Xe}}$.}
	\label{fig:distros}
\end{figure}
We numerically calculate the electron wavefunctions according to \cite{Kotila:2012zza} and as described in the previous section. Combining Eqs.~\eqref{eq:C}, \eqref{eq:w}, \eqref{eq:aterm}, and \eqref{eq:bterm}, we then determine the single electron distribution $d\Gamma/dE_1$ and the angular correlation $\alpha(E_1)$ for the three relevant phase space factors that can occur under the presence of short-range operators: $f_{11}^{(a)}$ (for mechanisms $i=1,2,3$ with a scalar electron current), $f_{66}^{(a)}$ (for mechanisms $i=4,5$ with an axial-vector electron current) and $f_{16}^{(0)}$ (for interference between the two classes). In the latter case, the contribution $f_{16}^{(1)}$ to the angular coefficient $b(E_1)$ vanishes identically. The electron phase space distribution $f_{11}$ is identical to that of the standard mass mechanism, calculated in the closure approximation. The results are shown in Fig.~\ref{fig:distros} for the $\ovbb$ decay isotopes $\prescript{76}{}{\text{Ge}}$, $\prescript{130}{}{\text{Te}}$ and $\prescript{136}{}{\text{Xe}}$. We plot both the normalized single energy distributions and the angular correlation as functions of the kinetic energy $E_1^\text{kin} = E_1 - m_e$ of one of the electrons, i.e. the range is from zero up to $Q_{\beta\beta}$ value of the respective isotope.

In all scenarios, the single energy distribution $d\Gamma/dE^\text{kin}_1$ is of a \emph{hill}-like shape, i.e. the two electrons preferably share the available kinetic energy equally. There is only a small difference between the $f_{11}$ and $f_{66}$ case, with the latter having a slightly flatter profile, that is unlikely to be distinguishable experimentally. The term $f_{16}$, corresponding to an interference between $i=1,2,3$ and $i=4,5$ mechanisms, has a more significantly flatter profile.

The angular correlation $\alpha(E^\text{kin}_1)$ is negative for $f_{11}$ and positive for $f_{66}$, i.e. in the former case, the electrons are preferably emitted back-to-back whereas in the latter case they preferably fly in a similar direction. This allows to potentially distinguish the scenarios resulting in $f_{66}$ from the standard mass mechanism as well as from scenarios corresponding to $f_{11}$.

Following \cite{Kotila:2012zza}, we define the integrated PSFs
\begin{align}
\label{eq:Gs}
	G^{(a)}_{ij} = \frac{2C}{\ln 2}\frac{g^{(a)}_{ij}}{4 R_A^2}
	               \int^{Q_{\beta\beta} + m_e}_{m_e} dE_1 
	                    w(E_1) f^{(a)}_{ij}(E_1,Q_{\beta\beta} + 2m_e - E_1),
\end{align}
with $g^{(0)}_{11} = 2$, $g^{(1)}_{11} = 2$, $g^{(0)}_{66} = 1/8$, $g^{(1)}_{66} = 1/8$, $g^{(0)}_{16} = 1$, $g^{(1)}_{16} = 0$ as well as $C$ and $w(E_1)$ as defined in Eqs.~\eqref{eq:C} and \eqref{eq:w}, respectively. The factor $1/R_A^2$ has been introduced in Eq.~\eqref{eq:Gs} to conform with standard notation to compensate for the corresponding factor in the NMEs as discussed above. The numerical values of the PSFs $G_{ij}^{(a)}$, calculated in analogy with \cite{Kotila:2012zza}, are given in Tab.~\ref{tab:psfs}. As mentioned before, the PSF $G_{16}^{(1)}$ vanishes identically; in addition, $G_{11-}^{(1)}$, corresponding to the interference between a right-handed and left-handed scalar electron current is also zero as indicated in Tab.~\ref{tab:psfs}.
\begin{table}[t!]
\centering
\begin{tabular}{c|rrr|rrrr|rr} \hline
	$G^{(i)}_{jk}$~[$10^{-15}~\text{y}^{-1}$] & 
	$G^{(0)}_{11}$ & $G^{(0)}_{66}$ & $G^{(0)}_{16}$ &
	$G^{(1)}_{11+}$ & $G^{(1)}_{11-}$ & $G^{(1)}_{66}$ & $G^{(1)}_{16}$ &
	$K_{11}$ & $K_{66}$ \\\hline
	$\prescript{76}{}{\text{Ge}}$  & 
	$4.72$ & $2.64$ & $1.7$ & $-3.90$ & $0$ & $1.95$ & 0 & $-0.83$ & $0.74$ \\
	$\prescript{82}{}{\text{Se}}$ &
	$20.4$ & $10.8$ & $5.9$ & $-18.1$ & $0$ & $9.10$ & 0 & $-0.89$ & $0.84$ \\
	$\prescript{100}{}{\text{Mo}}$ & 
	$31.8$ & $17.0$ & $8.4$ & $-28.6$ & $0$ & $14.2$ & 0 & $-0.90$ & $0.84$ \\
	$\prescript{130}{}{\text{Te}}$ & 
	$28.4$ & $15.3$ & $8.7$ & $-24.8$ & $0$ & $12.4$ & 0 & $-0.87$ & $0.81$ \\ 
	$\prescript{136}{}{\text{Xe}}$ & 
	$29.2$ & $15.7$ & $9.0$ & $-25.4$ & $0$ & $12.7$ & 0 & $-0.87$ & $0.81$ \\
	\hline
\end{tabular}
\caption{Phase space factors $G^{(i)}_{jk}$ for selected nuclei, calculated in analogy with \cite{Kotila:2012zza}, and corresponding angular correlation factors $K_{11}$ and $K_{66}$ ($K_{16}=0$ is always identically zero). All PSFs are given in units of $10^{-15}~\text{y}^{-1}$ as indicated.}
\label{tab:psfs}
\end{table}

With the above PSFs, the inverse $\ovbb$ decay half-life is then given by
\begin{align}
\label{eq:totalhalflife}
	T_{1/2}^{-1} = 
	  G_{11}^{(0)} \left|\sum_{I=1}^3 \epsilon_I \nme_I  \right|^2 
	+ G_{66}^{(0)} \left|\sum_{I=4}^5 \epsilon_I \nme_I  \right|^2 
    \mp G_{16}^{(0)} \text{Re}\left[\left(\sum_{I=1}^3 \epsilon_I \nme_I  \right)                
	\left(\sum_{I=4}^5 \epsilon_I \nme_I\right)^*\right].
\end{align}
Analogous to Eq.~\eqref{eq:aterm}, the sign in front of $G_{16}^{(0)}$ is negative (positive) if the chirality of the electron scalar current involved in the interference term is $R$ ($L$).

We can also calculate the integrated angular correlation factors as
\begin{align}
	K_{jk} = \frac{B}{A} = \frac{G_{jk}^{(1)}}{G_{jk}^{(0)}},
\end{align}
in the three different cases $jk=11$, $66$, $16$. The numerical values of $K_{11}$ and $K_{66}$ are also listed in Tab.~\ref{tab:psfs} whereas $K_{16} = 0$ is always identically zero. As already discussed, in view of the opposite sign for $11$ and $66$ in Tab.~\ref{tab:psfs}, an eventual measurement of the angular correlation will allow a discrimination of the two types of non-standard mechanisms.

\begin{table}[t!]
	\centering
	\begin{tabular}{cc|cccccccc}\hline
		& $T_{1/2}^{\text{exp}}~[y]$ & 
		$|\epsilon_1^{XX}|$ & $|\epsilon_1^{LR}|$ & 
		$|\epsilon_2^{XX}|$ & 
		$|\epsilon_3^{XX}|$ & $|\epsilon_3^{LR}|$ & 
		$|\epsilon_4^{XX,LR}|$ & 
		$|\epsilon_5^{XX}|$ & $|\epsilon_5^{RL, LR}|$ \\\hline
		$\prescript{76}{}{\text{Ge}}$  & $5.3\times 10^{25}$~\cite{Agostini:2013mzu} 
		& 1.5 & 1.5 & 190 & 110 & 220 & 250 & 60 & 50 \\
		$\prescript{130}{}{\text{Te}}$ & $2.8\times 10^{24}$~\cite{Martinez:2017zma} 
		& 3.5 & 3.4 & 420 & 240 & 490 & 550 & 140 & 120 \\
		$\prescript{136}{}{\text{Xe}}$ & $1.1\times 10^{26}$~\cite{Gando:2012zm} 
		& 0.57 & 0.57 & 84 & 50 & 110 & 110 & 23 & 19 \\
		\hline
	\end{tabular}
	\caption{Estimates of upper limits on the absolute values of the $\epsilon_I$ couplings (in units of $10^{-10}$) from current experimental bounds, assuming only one contribution is different from zero at a time. The chiralities of the involved quark currents are specified, as the corresponding bounds differ. The label $XX$ stands for the case when both chiralities are the same, i.e. $XX = RR$ or $XX = LL$. The experimental bounds at 90\% confidence level, reported by recent searches at KamLAND-Zen \cite{Gando:2012zm}, GERDA \cite{Agostini:2013mzu} and CUORE \cite{Martinez:2017zma} are included.}
	\label{tab:bounds}
\end{table}
\subsection{Bounds on Couplings}
In principle, a given underlying particle physics model may give rise to several contributions, and/or mixing among the corresponding Wilson coefficients will induce contributions through radiative effects from the scale of new physics, through the electroweak scale and down to the scale of QCD. The above formulae for the decay rate take into account all possible short-range contributions where the $\epsilon_I$ factors are to be understood to be effective at the QCD scale. To determine the numerical sensitivity to the $\epsilon_I$ factors, we here make the commonly considered simplifying assumption that only one term $\epsilon_I$ is different from zero and thus only one mechanism contributes at a time. 

The resulting upper limits on the $\epsilon_I$ factors are estimated in Tab.~\ref{tab:bounds}, based on the above calculation of the $\ovbb$ half-life and using the most stringent experimental bounds for the isotopes $\prescript{76}{}{\text{Ge}}$~\cite{Agostini:2013mzu}, $\prescript{130}{}{\text{Te}}$~\cite{Martinez:2017zma} and $\prescript{136}{}{\text{Xe}}$~\cite{Gando:2012zm}. In Tab.~\ref{tab:boundsFuture} we then show the estimated reach with respect to the effective couplings assuming a common future experimental sensitivity of $T^{\text{future}}_{1/2} = 10^{27}$~y including two additional potentially interesting isotopes, $\prescript{82}{}{\text{Se}}$ and $\prescript{100}{}{\text{Mo}}$. In both tables, upper limits on the absolute values $|\epsilon_i^{XY}|$ of the effective couplings are shown. Different chiralities of the quark currents in the operators lead to different bounds as indicated; here $\epsilon_i^{XX}$ denotes the case where the chiralities of the two quark currents are equal, i.e. $XX=LL$ and $XX=RR$. For $\epsilon_2$ and $\epsilon_4$, the bounds do not depend on the choice of chiralities.

\begin{table}[t!]
	\centering
	\begin{tabular}{cc|cccccccc}\hline
		& $T_{1/2}^{\text{exp}}~[y]$ & 
		$|\epsilon_1^{XX}|$ & $|\epsilon_1^{LR}|$ & 
		$|\epsilon_2^{XX}|$ & 
		$|\epsilon_3^{XX}|$ & $|\epsilon_3^{LR}|$ & 
		$|\epsilon_4^{XX,LR}|$ & 
		$|\epsilon_5^{XX}|$ & $|\epsilon_5^{RL, LR}|$ \\\hline
		$\prescript{76}{}{\text{Ge}}$  & $10^{27}$
		& 0.35 & 0.35 & 44 & 26 & 50 & 58 & 14 & 11 \\
		$\prescript{82}{}{\text{Se}}$ & $10^{27}$ 
		& 0.19 & 0.19 & 25 & 15 & 30 & 34 & 7.8 & 6.5 \\
		$\prescript{100}{}{\text{Mo}}$ & $10^{27}$ 
		& 0.2 & 0.2 & 16 & 8.9 & 16 & 20 & 8.7 & 6.5 \\
		$\prescript{130}{}{\text{Te}}$ & $10^{27}$ 
		& 0.18 & 0.18 & 22 & 13 & 26 & 29 & 7.5 & 6.1 \\
		$\prescript{136}{}{\text{Xe}}$ & $10^{27}$
		& 0.19 & 0.19 & 28 & 17 & 35 & 38 & 7.6 & 6.4 \\
		\hline
	\end{tabular}
	\caption{Sensitivity estimates on the absolute values of the $\epsilon_I$ couplings (in units of $10^{-10}$) from a prospective future experimental sensitivity $T^{\text{future}}_{1/2} = 10^{27}$~y, assuming only one contribution is different from zero at a time. The label $XX$ stands for the case when both chiralities are the same, i.e. $XX = RR$ or $XX = LL$.}
	\label{tab:boundsFuture}
\end{table}
We would like to stress that in constructing these tables, we have used the values of $\nme_F,\nme_{GT}$ and $\nme_T$ given in Table \ref{tab:nmes} and we estimate the values of the other NMEs involved by replacing $(\vecl{q}/m_p)^2 = 0.01$ as a rough average, and by neglecting the effect of differently-shaped $\vecl{q}^2$-dependence of form factors. We do not attempt to assess the resulting uncertainty and more accurate results based on the actual calculation of the additional NMEs will be reported in a future work.

\subsection{QCD Running of Couplings}
\label{sec:qcdrunning}
The above limits on the effective couplings of exotic short-range mechanisms have been so far implicitly assumed to apply at the QCD scale $\Lambda_\text{QCD}\approx 1$~GeV with one coupling set different from zero. Following \cite{Gonzalez:2015ady} it is possible to be more accurate and assume each of these couplings to exist at a certain new physics scale $\Lambda_\text{NP} \approx 1$~TeV and run it down to $\Lambda_\text{QCD}$, where the appropriate bound can be set employing the experimental limit on $\ovbb$ decay half-life. Consequently, the values obtained in this way can be compared with constraints derived from collider experiments.

The Renormalization Group Equation (RGE) for a set of coupled Wilson coefficients $\vecl{c} = (c_1, c_2, \dots, c_n)^T$ reads
\begin{align} 
\label{eq:rge}
	\frac{d\vecl{c}(\mu)}{d\log\mu} = \gamma^T\cdot\vecl{c}(\mu),
\end{align}
where $\gamma$ is the anomalous dimension matrix in the $\overline{\text{MS}}$-scheme. At one-loop order, it is given by $\gamma = -2(b - 2C_F\mathbb{I})$, with $C_F$ being the colour factor and $b$ a $\mu$-independent constant matrix. The solution to Eq.~\eqref{eq:rge} is most conveniently written in matrix form,
\begin{align}
	\vecl{c}(\mu) = U(\mu,\Lambda_\text{NP})\cdot\vecl{c}(\Lambda_\text{NP}),
\end{align}
where $\mu$ and $\Lambda_\text{NP}$ are the low and high energy scales, respectively, between which the coefficients are evolved.

We apply the procedure sketched above to the case of the effective couplings $c_I \equiv \epsilon_I(1~\text{TeV})$ of short-range operators triggering $\ovbb$ decay at the scale $\Lambda_\text{NP} = 1$~TeV. In this case, the evolution matrix $U = U(\Lambda_\text{QCD},\Lambda_\text{NP})$ of Wilson coefficients between $\Lambda_\text{NP}$ and $\Lambda_\text{QCD}$ is rather sparse; its only non-zero elements read \cite{Gonzalez:2015ady}
\begin{gather}
	U^{XX}_{(12)} =
	\begin{pmatrix}
  		 \phantom{-}2.39 & 0.02 \\
  		-3.83 & 0.35
	\end{pmatrix},\quad 
	U^{LR}_{(31)} =
	\begin{pmatrix}
		0.84 & -2.19 \\
		0    &  \phantom{-}4.13
	\end{pmatrix},\quad
	U^{XX}_{(45)} =
	\begin{pmatrix}
		 \phantom{-}0.35  & -0.96i \\
		-0.06i & 2.39
	\end{pmatrix},\nonumber\\
	U^{XX}_{(3)} = 0.70,\quad
	U^{LR}_{(4)} = 0.62,\quad
	U^{LR}_{(5)} = 4.13,
\label{eq:U-num}
\end{gather} 
where the subscripts denote the respective short-range operator(s) and the superscripts the chiralities of the quark currents involved. For example, the matrix $U^{XX}_{(12)}$ describes the mixing between the first and second short-range operators involving quark currents with the same chiralities. Using Eq.~\eqref{eq:U-num} and the approximated values of NMEs Eqs.~\eqref{eq:nme1}-\eqref{eq:nme5}, the corresponding bounds on couplings $c_I$ are obtained and shown in Tab.~\ref{tab:boundsQCD}. Analogous to Tab.~\ref{tab:bounds}, we take only one effective coupling different from zero at a time. The difference is that we make this assumption now at the scale $\Lambda_\text{NP}$ and we use the above Wilson RGE solution to evolve the couplings to $\Lambda_\text{QCD}$ to calculate the $\ovbb$ decay rate, potentially with more than one coupling active due to mixing. 

\begin{table}[t!]
\centering
\begin{tabular}{cc|ccccccccc}\hline
& $T_{1/2}^{\text{exp}}~[y]$ & 
$|c_1^{XX}|$ & $|c_1^{LR}|$ & 
$|c_2^{XX}|$ & 
$|c_3^{XX}|$ & $|c_3^{LR}|$ & 
$|c_4^{XX}|$ & $|c_4^{LR}|$ & 
$|c_5^{XX}|$ & $|c_5^{RL, LR}|$ \\\hline
$\prescript{76}{}{\text{Ge}}$ & $5.3\times 10^{25}$~\cite{Agostini:2013mzu} 
& 0.62 & 0.36 & 88 & 160 & 260 & 580 & 400 & 25 & 12 \\
$\prescript{130}{}{\text{Te}}$ & $2.8\times 10^{24}$~\cite{Martinez:2017zma} 
& 1.4 & 0.83 & 200 & 350 & 580 & 1300 & 880 & 59 & 28 \\
$\prescript{136}{}{\text{Xe}}$ & $1.1\times 10^{26}$~\cite{Gando:2012zm} 
& 0.24 & 0.14 & 32 & 72 & 130 & 250 & 190 & 9.6 & 4.7 \\
\hline
\end{tabular}
\caption{As Tab.~\ref{tab:bounds}, but for the effective couplings defined at the average new physics scale $\Lambda_\text{NP} = 1$~TeV.}
\label{tab:boundsQCD}
\end{table}

The resulting bounds on the couplings $|c_I|$ at $\Lambda_\text{NP}$, including QCD running effects, displayed in Tab.~\ref{tab:boundsQCD}, are weaker or more stringent than those in Tab.~\ref{tab:bounds}, depending on the operator in question. It should be noted that the limit on $|\epsilon_4|$ splits into two different values $c_4^{XX}$ and $c_4^{LR}$, since the running depends on the quark current chiralities. As for the effects of operator mixing, the limits on $|c_4|$ are for example less stringent due to the size of the corresponding RG evolution matrix elements, which are always smaller than $1$. In case of the mixing between $\mathcal{O}_1^{XX}$ and $\mathcal{O}_2^{XX}$, the limit on $|c^{XX}_2|$ is not much affected by the mixing because the relevant element of the evolution matrix is small, $[U^{XX}_{(12)}]_{12} = 0.02$; hence, the expectedly strong contribution from $\mathcal{O}_1^{XX}$ (large NMEs) to $c_2^{XX}$ is suppressed. As a result, the bounds in Tab.~\ref{tab:boundsQCD} are not drastically different from those in Tab.~\ref{tab:bounds}, despite the strong variation in sensitivity to the couplings $\epsilon_i$.

\section{Summary and Conclusion}
\label{sec:summary}

In this article we have developed a general formalism for short-range mechanisms contributing to neutrinoless double beta decay in an effective operator approach. Such contributions will arise when lepton number is broken at a new physics scale $\Lambda_\text{NP}$ much larger than the typical energy scale of $\ovbb$ decay $q \approx 100$~MeV. We have calculated the expected $\ovbb$ half lives by making use of the phase space factors calculated in the same way, as described in \cite{Kotila:2012zza} and nuclear matrix elements to leading order in $\vecl{q}/m_p$ of \cite{Barea:2015kwa}, where we especially elucidate the different contributions arising from general form factors in the nucleon currents. We also evaluate new phase space factors originating from the electron currents, including interference effects of different short-range contributions. The results of the present paper, complement those of Ali et al. \cite{Ali:2007ec} for long-range neutrinoless double beta decay. We also find that the angular correlation between the two emitted electrons is different for different mechanisms, although in our case, there are only two types of angular correlations, one for terms $1,2,3$ and one for terms $4,5$ of the effective Lagrangian.

Using experimental bounds on half-lives and estimating the novel matrix elements arising in short-range contributions, we have calculated the numerical limits on the effective new physics parameters $\epsilon_I$. To leading order, only the standard Fermi, Gamow-Teller matrix elements appear, but especially the enhanced values of the exotic and induced pseudo-scalar couplings in the form factor approach necessitate the inclusion of higher order terms in $\vecl{q}/m_p$. This then requires the determination of different nuclear matrix elements $\mathcal{M}_F^{\prime}$, $\mathcal{M}_{GT}^{\prime}$, $\mathcal{M}_T^{\prime}$, $\mathcal{M}_F^{\prime\prime}$, $\mathcal{M}_{GT}^{\prime\prime}$, $\mathcal{M}_T^{\prime\prime}$, the calculation of which is currently under way and will be presented in a subsequent publication. They are crucially important to accurately determine dominant contributions to short-range $\ovbb$ decay and to verify the strong limits we obtain on the effective new physics parameters ranging between $\epsilon_I \approx 10^{-10}$ to $10^{-7}$, which correspond to new physics scales in the multi-TeV region. Short-range contributions scale as $\epsilon \propto 1/\Lambda_\text{NP}^5$, and thus an increase in sensitivity on $\epsilon$ by an order of magnitude will only improve a limit on $\Lambda_\text{NP}$ by a factor of $\approx 1.6$. Nevertheless, accurate calculations of the limits and sensitivities are crucially important as they probe the phenomenologically interesting TeV scale. A robust map of potential sources of lepton number violation in this energy region will help us to get a better understanding of the mechanism of neutrino mass generation.

\textbf{Note:} While finalizing this manuscript, we noticed the preprint \cite{Cirigliano:2018yza}, which discusses short-range contributions as well, but using a complementary approach based on chiral effective field theory.

\section*{Acknowledgements}

This work was supported in part by the U.S. Department of Energy (Grant No. DE-FG-02-91ER-40608) and the UK Royal Society International Exchange program. The work of JK was supported by the V{\"a}is{\"a}l{\"a} Foundation and the Academy of Finland (Grant No. 314733). The work of LG and FFD was supported by a UK Science \& Technology Facilities Council Consolidated Grant. LG and FFD would like to thank Martin Hirsch for useful discussions and support on the QCD renormalization running, as well as Amr Hossameldin for carefully checking the nucleon current products.

\appendix
\section{Nucleon Current Products}
\label{sec:currentproducts}

In the following we explicitly show the products of the non-relativistically approximated hadronic currents for each of the five terms of the effective short-range Lagrangian Eq. \eqref{eq:lagsr}. Generally, we include all the terms up to linear order in $\vecl{q}/m_p$, from higher orders only terms enhanced by large form factors $F_{PS}$ and/or $F_{P}$ are kept. All the products are symmetrized in indices $a \leftrightarrow b$ labelling individual nucleons in the nuclei. In case of each term we keep track of signs corresponding to different combinations of chiralities and these signs we show as a row vector in front of every single term of the expressions. In case of the first three products of hadronic currents (i.e. those proportional to $\epsilon_1$, $\epsilon_2$ and $\epsilon_3$) three possibilities are presented and they correspond to the following combinations of chiralities (in this ordering): $RR$, $LL$ and $(1/2)\nba{RL + LR}$. For the fourth and fifth product (proportional to $\epsilon_4$ and $\epsilon_5$) a row of four signs is shown, as in those cases the two hadronic currents have different Lorentz structures, thus, all the four possible combinations of chiralities have to be considered (in this ordering): $RR$, $LL$, $RL$ and $LR$.

\paragraph*{Term 1: $J J j$} The product of currents $JJ$ is
\begin{align}
	\Pi_1 \equiv~& \frac{1}{2}\sqb{J_{\circ,a} J_{\circ,b} + J_{\circ,b} J_{\circ,a}} \\
	=~& \signs{+}{+}{+} F_S^2(q^2) I_a I_b \nn \tag*{$[\mathcal{O}(1)~S-S]$} \nn \\
	&\signs{+}{+}{-} \frac{F_{PS}^2(q^2)}{4m_p^2}(\vecs{\sigma}_a\cdot\vecl{q})
	     (\vecs{\sigma}_b\cdot\vecl{q}) \tag*{$[\mathcal{O}(100)~S-S]$} \nn
	+\dots, \nn
\end{align}
where the term proportional to $F_{PS}^2(q^2)$ can be re-coupled using the following relation
\begin{align}
	(\vecs{\sigma}_a\cdot\vecl{q}) (\vecs{\sigma}_b\cdot \vecl{q}) &=  
	\frac{1}{3} (\vecs{\sigma}_a\cdot\vecs{\sigma}_b)\vecl{q}^2
	-\frac{1}{3} \left[\vecl{q}^2-\frac{1}{3}(\vecl{q}\cdot\vecl{\op{r}}_{ab})^2\right] \vecl{S}_{ab},
\end{align}
with $\vecl{S}_{ab} = 3(\vecs{\sigma}_a \cdot \vecl{\op{r}}_{ab}) (\vecs{\sigma}_b \cdot \vecl{\op{r}}_{ab}) - (\vecs{\sigma}_a \cdot \vecs{\sigma}_b)$.

\paragraph*{Term 2: $J^{\mu\nu} J_{\mu\nu} j$}
For the second term of the short-range part of the Lagrangian we get the following approximation of the nuclear currents
\begin{align}
	\Pi_2 \equiv~& \frac{1}{2}\sqb{J^{\mu\nu}_{\circ,a} {J_{\circ \mu\nu,b}} + J^{\mu\nu}_{\circ,b} {J_{\circ \mu\nu,a}} } \\
	=~& \signs{-}{-}{-} 2F^2_{T_1}(q^2) \nb{\vecs{\sigma}_a\cdot\vecs{\sigma}_b} + \dots.
	\tag*{$[\mathcal{O}(1)~S-S]$}\nn
\end{align}

\paragraph*{Term 3: $J^{\mu} J_{\mu} j$} Approximating the nuclear currents for the third term we obtain
\begin{align}
	\Pi_3 \equiv~& \frac{1}{2} \sqb{ J^\mu_{\circ,a} J_{\circ \mu,b} + J^\mu_{\circ,b} J_{\circ \mu,a} } \\
	=~& \signs{+}{+}{+}F_V^2(q^2) I_a I_b
	\tag*{$[\mathcal{O}(1)~S-S]$}\nn \\
	& \signs{-}{-}{+} F_A^2(q^2) (\vecs{\sigma}_a\cdot\vecs{\sigma}_b)
	\tag*{$[\mathcal{O}(1)~S-S]$}\nn \\
	&\signs{+}{+}{-} 2\frac{F_A(q^2) F_P(q^2)}{4 m_p^2} (\vecs{\sigma}_a\cdot\vecl{q})(\vecs{\sigma}_b\cdot\vecl{q}) 
	\tag*{$[\mathcal{O}(1)~S-S]$}\nn \\
	&\signs{+}{+}{+} \frac{\nba{F_V(q^2)+F_{W}(q^2)}^2}{4m_p^2}(\vecs{\sigma}_a\times\vecl{q})(\vecs{\sigma}_b\times\vecl{q}) 
	\tag*{$[\mathcal{O}(0.1)~S-S]$} \nn \\
	&\signs{-}{-}{+} \frac{F_P^2(q^2)}{16m_p^4}\vecl{q}^2 
		(\vecs{\sigma}_a\cdot\vecl{q})(\vecs{\sigma}_b \cdot \vecl{q}) 
		\tag*{$[\mathcal{O}(1)~S-S]$}
	+ \dots. \nn
\end{align}
where the term proportional to $\nba{F_V(q^2)+F_{W}(q^2)}^2$ can be re-coupled as follows
\begin{align}
	(\vecs{\sigma}_a\times \vecl{q}) (\vecs{\sigma}_b\times \vecl{q}) &=  
	-\frac{1}{3} (\vecs{\sigma}_a\cdot\vecs{\sigma}_b)\vecl{q}^2 - \frac{1}{6} \left[\vecl{q}^2-\frac{1}{3}(\vecl{q}\cdot\vecl{\op{r}}_{ab})^2\right] \vecl{S}_{ab},
\end{align}
and as before $\vecl{S}_{ab} = 3(\vecs{\sigma}_a \cdot \vecl{\op{r}}_{ab}) (\vecs{\sigma}_b \cdot \vecl{\op{r}}_{ab}) - (\vecs{\sigma}_a \cdot \vecs{\sigma}_b)$.

\paragraph*{Term 4: $J^\mu J_{\mu\nu} j^\nu$}
The product of tensor and vector nuclear current in the fourth term can be non-relativistically approximated as
\begin{align}
\Pi_{4\nu,ab} &\equiv \frac{1}{2}\sqb{J^{\mu}_{\circ,a} J_{\circ\mu\nu,b} + J^{\mu}_{\circ,b} J_{\circ\mu\nu,a} } \\
		&\approx {g_{\nu}}^{0}\Bigg\{ \foursigns{-}{-}{+}{+} iF_A(q^2)F_{T_1}(q^2)\nba{\vecs{\sigma}_a\cdot\vecs{\sigma}_b} 
		\tag*{$[\mathcal{O}(1)~S-S]$} \nn \\
		&\hspace{1.45cm}\foursigns{+}{+}{-}{-} i\frac{F_P(q^2)F_{T_1}(q^2)}{4 m_p^2} \nba{\vecs{\sigma}_a\cdot\vecl{q}} \nba{\vecs{\sigma}_b\cdot\vecl{q}}
		\tag*{$[\mathcal{O}(1)~S-S]$} \nn \Bigg\} \\
		&\hspace{0.27cm}+ {g_{\nu}}^{i} \Bigg\{
		\foursigns{+}{-}{-}{+} \frac{i}{2} F_V(q^2)F_{T_1}(q^2) \nba{ I_a\sigma_{b i} + I_b\sigma_{a i} } 
		\tag*{$[\mathcal{O}(1)~S-P]$} \nn \\
		&\hspace{1.45cm}\foursigns{-}{-}{-}{-} i\frac{F_V(q^2)\sqb{F_{T_1}(q^2)-2F_{T_2}(q^2)}}{2m_p}q_iI_aI_b
		\tag*{$[\mathcal{O}(0.1)~S-P]$} \nn \\
		&\hspace{1.45cm}\foursigns{-}{-}{-}{-} \frac{F_V(q^2)F_{T_1}(q^2)}{4m_p} \nn \\
		& \hspace{1.45cm} \times \sqb{ I_a(\vecs{\sigma}_b\times\vecl{Q})_i + I_b(\vecs{\sigma}_a\times\vecl{Q})_i } 
		\tag*{$[\mathcal{O}(0.1)~S-P]$} \nn \\
		&\hspace{1.45cm}\foursigns{-}{-}{-}{-} \frac{F_V(q^2)F_{T_1}(q^2)}{4m_p} \times \sqb{ I_a (\vecs{\sigma}_b\times\vecl{Q})_i + I_b (\vecs{\sigma}_a\times\vecl{Q})_i } 
		\tag*{$[\mathcal{O}(0.1)~S-P]$} \nn \\
		&\hspace{1.45cm}\foursigns{-}{-}{-}{-} i \frac{\sqb{F_V(q^2)+F_W(q^2)}F_{T_1}(q^2)}{4m_p} \nn \\
		& \hspace{1.45cm} \times \sqb{ 2 q_i(\vecs{\sigma}_a\cdot\vecs{\sigma}_b) - \sigma_{ai}(\vecl{q}\cdot\vecs{\sigma}_b) - \sigma_{bi}(\vecl{q} \cdot\vecs{\sigma}_a) }
		\tag*{$[\mathcal{O}(0.1)~S-P]$} \nn \\
		&\hspace{1.45cm}\foursigns{-}{-}{+}{+} \frac{F_A(q^2)F_{T_1}(q^2)}{4m_p} \sqb{ (\vecs{\sigma}_a\cdot\vecl{Q})\sigma_{b i} + (\vecs{\sigma}_b\cdot\vecl{Q})\sigma_{a i} } 
		\tag*{$[\mathcal{O}(0.1)~S-P]$} \nn \\
		&\hspace{1.45cm}\foursigns{+}{+}{-}{-} \frac{F_A(q^2)\sqb{F_{T_1}(q^2)-2F_{T_2}(q^2)}}{4m_p} \nn \\
		& \hspace{1.45cm} \times \sqb{ (\vecs{\sigma}_a\times\vecl{q})_i I_b + (\vecs{\sigma}_b\times\vecl{q})_i I_a } 
		\tag*{$[\mathcal{O}(0.1)~S-P]$} \nn \\
		&\hspace{1.45cm}\foursigns{-}{-}{+}{+} i \frac{F_A(q^2) F_{T_1}(q^2)}{4m_p} 
		 \nn \\
		& \hspace{1.45cm} \times \sqb{ \sigma_{ai}(\vecl{Q}\cdot\vecs{\sigma}_b) + \sigma_{bi}(\vecl{Q} \cdot\vecs{\sigma}_a) - 2 Q_i(\vecs{\sigma}_a\cdot\vecs{\sigma}_b) }
		\tag*{$[\mathcal{O}(0.1)~S-P]$} \nn \\
		&\hspace{1.45cm}\foursigns{-}{-}{+}{+} i \frac{F_P(q^2) F_{T_1}(q^2)}{8 m_p^2} q^0 \times \sqb{ (\vecs{\sigma}_a\cdot\vecl{q})\sigma_{b i} + (\vecs{\sigma}_b\cdot\vecl{q})\sigma_{a i} } 
		\tag*{$[\mathcal{O}(0.1)~S-P]$} \nn \\
		&\hspace{1.45cm}\foursigns{+}{+}{-}{-} \frac{F_P(q^2) F_{T_1}(q^2)}{16 m_p^3} 
		\sqb{ \sigma_{ai}(\vecl{q} \cdot\vecl{Q})(\vecl{q} \cdot\vecs{\sigma}_b) \right. \nn \\
		&\left. \hspace{1.45cm} + \sigma_{bi}(\vecl{q} \cdot\vecl{Q})(\vecl{q} \cdot\vecs{\sigma}_a) - 2 Q_i(\vecl{q}\cdot\vecs{\sigma}_b)(\vecl{q}\cdot\vecs{\sigma}_b) }
		\tag*{$[\mathcal{O}(0.1)~S-P]$} \nn
		\Bigg\} + \dots. \nn
\end{align}

\vfill

\paragraph*{Term 5: $J^\mu J j_\mu$} Approximating the nuclear currents in this case we obtain
\begin{align}
\Pi^\mu_5 &\equiv 
	\frac{1}{2}\sqb{J^\mu_{\circ,a} J_{\circ,b} + J^\mu_{\circ,b} J_{\circ,a}} \\
		&\approx {g^{\mu}}_{0}\Bigg\{\foursigns{+}{+}{+}{+} F_S(q^2)F_V(q^2)I_a I_b 
		\tag*{$[\mathcal{O}(1)~S-S]$} \nn \\
		&\hspace{1.5cm}\foursigns{+}{+}{-}{-} \frac{F_{PS}(q^2) F_A(q^2)}{8m_p^2} \sqb{ (\vecs{\sigma}_a\cdot\vecl{Q})(\vecs{\sigma}_b\cdot\vecl{q}) + (\vecs{\sigma}_a\cdot\vecl{q})(\vecs{\sigma}_b\cdot\vecl{Q}) }
		\tag*{$[\mathcal{O}(1)~S-S]$} \nn \\
		&\hspace{1.5cm}\foursigns{-}{-}{+}{+} \frac{F_{PS}(q^2) F_P(q^2)}{8m_p^3} q^0 (\vecs{\sigma}_a\cdot\vecl{q})(\vecs{\sigma}_b\cdot\vecl{q}) \Bigg\}
		\tag*{$[\mathcal{O}(1)~S-S]$} \nn \\
		&\hspace{0.2cm}+ {g^{\mu}}_{i}\Bigg\{ \foursigns{-}{+}{-}{+} \frac{F_S(q^2) F_A(q^2)}{2} \nba{\sigma_a^i I_b + \sigma_b^i I_a} 
		\tag*{$[\mathcal{O}(1)~S-P]$} \nn \\
		&\hspace{1.5cm}\foursigns{-}{-}{-}{-} \frac{F_S(q^2) F_V(q^2)}{2m_p} Q^i I_a I_b
		\tag*{$[\mathcal{O}(0.1)~S-P]$} \nn \\
		&\hspace{1.5cm}\foursigns{+}{+}{+}{+} i \frac{F_S(q^2) \sqb{F_V(q^2) + F_W(q^2)}}{4m_p}  \nn \\
		& \hspace{1.45cm} \times \sqb{ (\vecs{\sigma}_a\times\vecl{q})^i I_b + (\vecs{\sigma}_b\times\vecl{q})^i I_a }
		\tag*{$[\mathcal{O}(0.1)~S-P]$} \nn \\
		&\hspace{1.5cm}\foursigns{-}{-}{+}{+} \frac{F_{PS}(q^2) F_A(q^2)}{4m_p} \sqb{ \sigma_a^i(\vecs{\sigma}_b\cdot\vecl{q}) + \sigma_b^i(\vecs{\sigma}_a\cdot\vecl{q}) }
		\tag*{$[\mathcal{O}(10)~S-P]$} \nn \\
		&\hspace{1.5cm}\foursigns{-}{+}{+}{-} \frac{F_{PS}(q^2) F_V(q^2)}{8m_p^2} Q^i \sqb{ I_a (\vecs{\sigma}_b\cdot\vecl{q}) + I_b (\vecs{\sigma}_a\cdot\vecl{q}) } 
		\tag*{$[\mathcal{O}(1)~S-P]$} \nn \\
		&\hspace{1.5cm}\foursigns{+}{-}{+}{-} \frac{F_{S}(q^2) F_P(q^2)}{8m_p^2} q^i \sqb{ (\vecs{\sigma}_a\cdot\vecl{q}) I_b + (\vecs{\sigma}_b\cdot\vecl{q}) I_a}
		\tag*{$[\mathcal{O}(1)~S-P]$} \nn \\
		&\hspace{1.5cm}\foursigns{+}{+}{-}{-} \frac{F_{PS}(q^2) F_P(q^2)}{8m_p^3} q^i (\vecs{\sigma}_a\cdot\vecl{q}) (\vecs{\sigma}_b\cdot\vecl{q})
		\tag*{$[\mathcal{O}(10)~S-P]$}
		\Bigg\} + \dots. \nn
\end{align}

\vfill

\section{Leptonic Matrix Elements}
\label{sec:leptonicmatrixelements}

\paragraph*{Terms 1, 2 and 3}
The electron current for these terms is
\begin{equation}
	j=\bar{e_1}(x)(1 \pm \gamma_5) e_2^C(x).
\end{equation}
Note that both the electron wavefunctions depend on the same coordinate variable, as a contact interaction is considered. In the $S_{1/2}-S_{1/2}$ wave approximation and using Tomoda's notation we obtain
\begin{align}
	\bar{e_1}(1 \pm \gamma_5) e_2^c &\approx 
	(\bar{e}_{\vecl{p}_1 s})^{S_{1/2}}(1\pm\gamma_5)(e^{C}_{\vecl{p}_2 s'})^{S_{1/2}} 
	\nn \\
	&= \nba{e_{\vecl{p}_1 s}^{S_{1/2}}}^\dagger \gamma_0 (1\pm\gamma_5) i \gamma_2 \nba{e_{\vecl{p}_2 s'}^{S_{1/2}}}^* \nn \\
	&= \begin{pmatrix} 
			g_{-1}(\epsilon_1, r) \chi_s^{\dagger} & 
			f_{1}(\epsilon_1, r)\chi_s^\dagger(\boldsymbol{\sigma}\cdot\vecl{\op{p}}_1) 
		\end{pmatrix} 
		\gamma_0 \left( 1 \pm \gamma_5\right) i \gamma_2 
		\begin{pmatrix}
			g_{-1} (\epsilon_2, r) \chi_{s'} \\ f_{1} (\epsilon_2, r)(\boldsymbol{\sigma}\cdot\vecl{\op{p}}_2) \chi_{s'}
		\end{pmatrix}, 
\label{eq:PSFder1}
\end{align}
where all the matrices are considered in the standard Dirac representation,
\begin{align}
	\gamma_0 = 
	\Bigg(\begin{array}{cc}
		1 & 0 \\
		0 & -1
	\end{array} \Bigg), \quad
	\vecs{\gamma} = 
	\Bigg( \begin{array}{cc}
		0 & \vecs{\sigma} \\
		-\vecs{\sigma} & 0
	\end{array} \Bigg), \quad
	\gamma_5 = \Bigg( \begin{array}{cc}
		0 & 1 \\
		1 & 0
	\end{array} \Bigg), \quad
	C = i\gamma_2\gamma_0 = 
	\Bigg( \begin{array}{cc}
		0 & -i \sigma_2 \\
		-i \sigma_2 & 0
	\end{array} \Bigg).
\end{align}
Next, we expand and square Eq.~\eqref{eq:PSFder1}. After summing over spins and using the properties of the spinors $\chi_s$ we get
\begin{align}
&\sum_{s,s'}
\left[\left( {f^{-1}}_{1} \chi^\dagger_s (\vecs{\sigma} \cdot \vecl{\op{p}}_2) \sigma_2 \chi_{s'} + {f_1}^{-1} \chi_s^\dagger (\vecs{\sigma} \cdot \vecl{\op{p}}_1) \sigma_2 \chi_{s'} \right) \right. \nn\\
&\pm  \left. \left(f^{-1-1}\chi^\dagger_s \sigma_2 \chi_{s'}+f_{11}\chi^\dagger_s (\vecs{\sigma} \cdot \vecl{\op{p}}_1) (\vecs{\sigma} \cdot \vecl{\op{p}}_2) \sigma_2 \chi_{s'} \right) \right]^2 \nn\\
&= 2 \left[ f_{11}^{(0)} + f_{11+}^{(1)} (\vecl{\op{p}}_1 \cdot \vecl{\op{p}}_2) \right], 
\end{align}
where $f_{11}^{(0)}$ and $f_{11+}^{(1)}$ are defined in Eq.~\eqref{eq:PSFscal}. For the interference term combining a left-handed with a right-handed electron current, the calculation is analogous to the procedure shown above, but the plus sign in the definition of $f^{(1)}_{11}$ would change to a minus sign.

\paragraph*{Terms 4 and 5}
The electron current for these terms is
\begin{equation}
	j^{\mu} \equiv \bar{e_1}(x) \gamma^{\mu} \gamma_5 e^C_2(x).
\end{equation}

\pagebreak

In $S_{1/2}-S_{1/2}$ approximation we have
\begin{align}
	\bar{e_1} \gamma^{\mu} \gamma_5 e^C_2 &\approx 
	(\bar{e}_{\vecl{p}_1 s})^{S_{1/2}}\gamma_{\mu}
	\gamma_5(e^{C}_{\vecl{p}_2 s'})^{S_{1/2}} \nn\\
	&= \nba{e_{\vecl{p}_1 s}^{S_{1/2}}}^{\dagger} \gamma_0 \gamma_{\mu} \gamma_5 i \gamma_2 \nba{ e_{\vecl{p}_2 s'}^{S_{1/2}}}^* \nn \\
	&= \begin{pmatrix} g_{-1}(\epsilon_1, r) \chi_s^{\dagger} & f_{1}(\epsilon_1, r)\chi_s^{\dagger}(\boldsymbol{\sigma}\cdot\vecl{\op{p}}_1) \end{pmatrix} \gamma_0 \gamma_{\mu} \gamma_5 i \gamma_2 \begin{pmatrix}g_{-1} (\epsilon_2, r) \chi_{s'} \\ f_{1} (\epsilon_2, r)(\boldsymbol{\sigma}\cdot\vecl{\op{p}}_2) \chi_{s'} \end{pmatrix}.
\end{align}
For $\mu=0$, after squaring, summing over spins and using the properties of the spinors $\chi_s$, we obtain
\begin{align}
	\sum_{s,s'} \sqb{f^{-1-1}\chi_s^{\dagger}\sigma_2\chi_{s'} + f_{11}\chi_s^{\dagger}(\boldsymbol{\sigma}\cdot\vecl{\op{p}}_1)(\boldsymbol{\sigma}\cdot\vecl{\op{p}}_2)\sigma_2\chi_{s'}}^2
	= \frac{1}{8} \left[ f^{(0)}_{66} + f^{(1)}_{66}(\vecl{\op{p}}_1\cdot\vecl{\op{p}}_2) \right],
\end{align}
where $f_{66}^{(0)}$ and $f_{66}^{(1)}$ are defined in Eq.~\eqref{eq:PSFcal2}. A similar derivation is possible for spatial $\mu=k$; however, as stated in the main text, it does not enter the contributions to $0^+ \to 0^+$ transition.

\paragraph*{Interference between terms 1,2,3 - 4,5} For the interference between terms $1,2,3$ and terms $4,5$ we use the same procedure as before and for $\mu=0$ we obtain 
\begin{align}
	\sum_{s,s'}&\sqb{f^{-1-1}\chi_s^{\dagger}\sigma_2\chi_{s'} + f_{11}\chi_s^{\dagger}(\boldsymbol{\sigma}\cdot\vecl{\op{p}}_1)(\boldsymbol{\sigma}\cdot\vecl{\op{p}}_2)\sigma_2\chi_{s'}}^{\dagger} \nn\\
	\times &\left[\left( {f^{-1}}_{1} \chi^{\dagger}_s (\vecs{\sigma} \cdot \vecl{\op{p}}_2) \sigma_2 \chi_{s'} + {f_1}^{-1} \chi_s^{\dagger} (\vecs{\sigma}\cdot \vecl{\op{p}}_1) \sigma_2 \chi_{s'} \right) \right. \nn \\
	&\pm \left. \left(f^{-1-1}\chi^{\dagger}_s \sigma_2 \chi_{s'} 
	+ f_{11}\chi^{\dagger}_s (\vecs{\sigma} \cdot \vecl{\op{p}}_1) (\vecs{\sigma} \cdot \vecl{\op{p}}_2) \sigma_2 \chi_{s'} \right) \right] \nn \\
	=&\mp \frac{1}{2}\left[f_{16}^{(0)} 
	+ f_{16}^{(1)}(\vecl{\op{p}}_1\cdot\vecl{\op{p}}_2)\right],
\end{align}
where $f_{16}^{(0)}$ and $f_{16}^{(1)}$ are defined in Eq.~\eqref{eq:PSFcal3}. As before, we do not present the phase space factor for spatial $\mu = k$ as it does not enter the calculation of $0^+ \to 0^+$ transitions.

\bibliography{literature}
\bibliographystyle{h-physrev4}
\end{document}